\documentclass[12pt]{iopart}

\usepackage{graphicx}
\usepackage{tikz}
\usepackage{amssymb}
\usepackage{array}
\usepackage{lipsum}
\usepackage{url}
\usepackage{pgfplots}
\pgfplotsset{compat=1.16}
\usepackage{iopams}
\usepackage{hyperref}

\hypersetup{ colorlinks=true, linktoc=all, linkcolor=blue, }

\newcommand{\sqrts}{\sqrt{s_\textrm{\scriptsize NN}}}
\newcommand{\yexp}{y_{\textrm{\scriptsize exp}} }
\newcommand{\yexpj}{y_{\textrm{\scriptsize{exp},j}} }
\newcommand{\yexpk}{y_{\textrm{\scriptsize{exp},k}} }

\newcommand{\sigmaexpj}{\sigma_{\textrm{\scriptsize{exp},j}} }

\newcommand{\ythj}{y_{\textrm{\scriptsize{th},j}} }
\newcommand{\ythzero}{y_{\textrm{\scriptsize{th},0}} }
\newcommand{\ythk}{y_{\textrm{\scriptsize{th},k}} }

\newcommand{\sigmathj}{\sigma_{\textrm{\scriptsize{th},j}} }

\newcommand{\tauhydro}{\tau_{\textrm{\scriptsize hydro}}}

\begin{document}
	
	\title{Applications of emulation and Bayesian methods in heavy-ion physics}
	
	\author{Jean-Fran\c{c}ois Paquet}
	
	\address{Department of Physics and Astronomy and Department of Mathematics, Vanderbilt University, Nashville TN 37235}
	\ead{jean-francois.paquet@vanderbilt.edu}
	\vspace{10pt}
	\begin{indented}
		\item[]October 2023
	\end{indented}
	
	\begin{abstract}
		Heavy-ion collisions provide a window into the properties of many-body systems of deconfined quarks and gluons. Understanding the collective properties of quarks and gluons is possible by comparing models of heavy-ion collisions to measurements of the distribution of particles produced at the end of the collisions. 
		These model-to-data comparisons are extremely challenging, however, because of the complexity of the models, the large amount of experimental data, and their uncertainties.
		Bayesian inference provides a rigorous statistical framework to constrain the properties of nuclear matter by systematically comparing models and measurements.
		
		This review covers model emulation and Bayesian methods as applied to model-to-data comparisons in heavy-ion collisions. Replacing the model outputs (observables) with Gaussian process emulators is key to the Bayesian approach currently used in the field, and both current uses of emulators and related recent developments are reviewed. The general principles of Bayesian inference are then discussed along with other Bayesian methods, followed by a systematic comparison of seven recent Bayesian analyses that studied quark-gluon plasma properties, such as the shear and bulk viscosities. The latter comparison is used to illustrate sources of differences in analyses, and what it can teach us for future studies.
	\end{abstract}

	\newpage
	
	\tableofcontents
	
	\newpage
	
	\section{Introduction}
	
	The Relativistic Heavy Ion Collider and the Large Hadron Collider can accelerate and collide large nuclei at velocities close to the speed of light.
	The amount of energy concentrated at the point of impact is so large that it compares to the energy density of the Universe microseconds after the Big Bang.
	At this density, the protons, neutrons and nuclei that form ordinary nuclear matter melt into their constituent quarks and gluons.
	This microscopic hot and dense droplet of nuclear matter dissipates in an instant in the vacuum that surrounds it, exploding in a shower of subatomic particles.
	
	These collisions of heavy nuclei (heavy ions) provide a unique opportunity to study the properties of dense many-body systems of quarks and gluons, and their transition from and to ordinary nuclear matter. This information is only accessible indirectly, however: less than $10^{-22}$~second elapses between the impact of the nuclei, the release and interaction of quarks and gluons, and their subsequent recombination into bound states. The complex history of the collision must be reconstructed from the final shower of particles, which are measured by the multipurpose detectors that envelop the collision region. Relating properties of nuclear matter to the final shower of particles is challenging, and heavy-ion physics is a prime candidate to apply advanced inference techniques in model-to-data comparisons. This is the focus of the present review.
	
	\subsection{Parameter estimation in heavy-ion physics}
	
	From a theoretical point of view, it is a considerable undertaking to describe nucleus collisions from the initial impact of the nuclei to the final shower of particles observed by detectors. The problem is broadly approached piecewise. Nuclear structure properties of heavy nuclei, measured in scattering experiments, can be combined with effective field theory calculations to understand the state of the nuclei shortly after their impact.
	Many-body quantum field theory is used to study the interaction of energetic quarks and gluons with their lower-energy counterparts.
	In certain regimes, the collective motion of strongly-interacting quarks and gluons is described with relativistic fluid dynamics, using an equation of state calculated numerically from lattice quantum field theory. By combining these different theoretical approaches and more, one can obtain a multi-stage model that is a functional description of the successive stages of ultrarelativistic nuclear collisions. Any quantity that is (i) unknown, (ii) uncertain, or (iii) only partly constrained by theoretical guidance can be considered a model parameter that enters the description of the collision.
	For example, the shear and bulk viscosities of strongly-coupled nuclear matter are generally parametrized as a function of temperature, and the same can be done for the equation of state~\cite{Pratt:2015zsa} (particularly in regions of the phase diagram where it is not constrained well from lattice calculations).
	The term ``parameter estimation'' can be used for studies that attempt to constrain properties of nuclear matter by comparing a model of the collision to measurements.
	
	A large amount of data is available to constrain the model's parameters, a result of the comprehensive heavy-ion-collision program at the Relativistic Heavy Ion Collider and the Large Hadron Collider.
	The detectors measure the number and momentum distributions of different species of particles produced after each collision, as well as various correlations between these particles. 
	The final shower of particles fluctuates from one collision to the next, due to the variable degree of overlap of nuclei --- which cannot be controlled --- and the quantum nature of nuclear collisions. Measurements are defined by averaging over a large number of collisions.
	The variable degree of overlap of the nuclei --- the centrality --- provides a natural parameter that changes the system's size, geometry and density. By further colliding different nuclei species and varying the collision energy of the nuclei, one can probe the system in a large number of different configurations.
	
	From a parameter estimation point of view, the situation can thus be broadly described as follows. The number of measurements available is large, and the type of measurements varies from simple averages (e.g., particle numbers) to intricate correlation observables. The measurements are stochastic, with statistical uncertainties ranging from a fraction of a percent to tens of percent. There are also systematic uncertainties that can be correlated across different data sets, although information about these correlations is often not readily available.
	On the theory side, multistage models provide a general description of the collisions to compare with these measurements. There is uncertainty in the physics description of the collisions, represented in the literature by different models used to describe the same stage of the collision.
	For example, a given stage of the collision can be parametrized with a flexible ansatz~\cite{Moreland:2014oya,Ke:2016jrd,Soeder:2023vdn}, or one can use an ab initio model with fewer parameters but additional physics assumptions~\cite{Dumitru:2012yr,Schenke:2012wb,Schenke:2012fw,Gelis:2016upa}.
	The number of parameters can vary significantly, depending on modelling choices and assumptions.
	In practice, most systematic model-to-data comparisons use 5 to 20 model parameters.
	
	Measurements are not known to arbitrary precision, and parameters cannot be expected to be constrained to arbitrary precision either.
	An ensemble of values for the parameters will be consistent to various degrees with the data and their uncertainty, leading to a distribution of acceptable model parameters.
	Because most models are highly non-linear, there is a complex mapping between the measurements and the corresponding constraints on the parameters.
	Bayesian inference provides a systematic approach to performing this mapping, naturally incorporating existing constraints on the parameters --- from theory or previous experimental comparisons --- through the use of priors. It can account for experimental uncertainties as well as theoretical ones (for example, statistical uncertainties from stochastic simulations); correlations between observables are straightforward to include as well.
	Bayesian inference can be applied to problems with a large number of parameters and measurements, making it well-suited for studying complex systems like heavy-ion collisions.

	\subsection{Hard and soft physics, and observable factorization in heavy-ion collisions}
	
	Measurements tend to probe different features of the collisions. Put another way, measurements are sensitive to different aspects of the theoretical description of the collisions. 
	The most common example in heavy-ion collisions is the division between a ``hard'' and a ``soft'' sector. Broadly speaking, the soft sector refers to the lower-energy particles found in the final shower of particles, while ``hard sector'' is used for the higher-energy or higher-mass particles. The higher the energy or mass of a particle, the fewer are produced, and some very high-energy or very heavy particles may only be produced once every hundreds or thousands of collisions. These rare particles thus have a limited impact on observables that measure the average number of particles produced per collision, for example, or other ``soft sector observables''. Correspondingly, the physical processes that are relevant to describe these rare particles tend to be different from those that drive the soft sector physics: the hard sector generally allows for applications of perturbative theory, while non-perturbative physics typically dominates the soft sector. Bayesian inference of relativistic nuclear collisions is still largely divided along this soft/hard separation. The main examples used in this review will be from the soft sector; however,  as far as Bayesian inference is concerned, both the hard and soft sectors can be treated identically.

	\subsection{Structure of this review}
	
	This review focuses on emulation and Bayesian methods as applied to the study of heavy-ion collisions. Emulation can be seen as a practical numerical matter: when significant computational resources are necessary to simulate heavy-ion collisions, there is value in building a proxy for the model that can approximate the model's output for a broad range of model parameters. On the other hand, emulators can provide valuable physics insights, for example, to help visualize the effect of parameters on observables.
	
	In this review, we will take a close look at Bayesian studies of shear and bulk viscosities of nuclear matter. 
	It is by no means the only quantity of interest, but many important features and challenges of Bayesian inference can be discussed with this example.
	The same principles and ideas apply to the study of any other nuclear property, be it the equation of state of nuclear matter at finite baryon density or the interaction rate of high and low energy quarks and gluons (``parton energy loss'').
	
	We first provide a broad overview of Bayesian inference in heavy-ion collisions in Section~\ref{sec:brief_intro}. We next discuss emulation in Section~\ref{sec:emulation}, both the practical aspects, recent progress, and some relevant applications of emulators in studying heavy-ion collisions. We then discuss the different features of Bayesian inference in Section~\ref{sec:bayes}, leaving specific applications to heavy-ion data for Section~\ref{sec:bayes_applications}.
	Considerations for the future of Bayesian inference applications in heavy-ion physics are discussed in Section~\ref{sec:future}, followed by a summary in Section~\ref{sec:summary}.

	\section{Brief overview of emulation and Bayesian inference in heavy-ion collisions}
	
	\label{sec:brief_intro}
	
	The broad picture of model-to-data comparison in heavy-ion collisions was overviewed in the introduction. In this section, we provide additional details on the data, the model, its parameters, and the application of Bayesian inference to compare them. We use this section to provide context for the latter sections of the paper.
	
	\subsection{Model and parameters}
	
	While different approaches have been explored to study heavy-ion collisions, what is generally referred to as the ``standard model of heavy-ion collisions'' is a multi-stage model with relativistic viscous hydrodynamics at its core~\cite{Heinz:2013th, Gale:2013da, deSouza:2015ena,Busza:2018rrf}. Broadly speaking, the hydrodynamic simulation is the intermediate stage of the overall model, and it is convenient to divide the collision into a ``pre-hydrodynamics'', a hydrodynamic and a ``post-hydrodynamic'' stage. The pre-hydrodynamic stage effectively provides initial conditions for the hydrodynamic simulation. The post-hydrodynamic stage interfaces the hydrodynamic simulation and the final shower of particles measured by detectors. Observables are computed from the ensemble of particles found at the end of the post-hydrodynamic stage.
	
	Even within this approach, there is not one ``standard'' model of heavy-ion collisions but a family of variations. For example, the pre-hydrodynamic stage can rely on a flexible ansatz, or an ab initio calculation with fewer parameters but a set of physics assumptions and simplifications. It is not necessarily clear a priori which approach is superior. Even in terms of their ability to describe measurements, it can be difficult to declare a victor, given that one scenario generally has several more parameters than the other, as discussed in Section~\ref{sec:model_selection}.
	
	The multi-stage model has an ensemble of parameters whose physical significance varies. For example, the shear and the bulk viscosities are properties of nuclear plasmas and enter the relativistic viscous hydrodynamic simulation. The viscosities are functions of the temperature of the plasma, and can be parameterized in different ways (discussed in Section~\ref{sec:bayes_applications} and Table~\ref{table:visc_param}). In this case, one could say that the viscosities are the physical quantity, and the parameters attempt to approximate the temperature dependence of this physical quantity. Another common parameter is the time at which the hydrodynamic simulation is initialized, ``$\tauhydro$''. It is not a physical parameter in the sense that there should be a smooth transition between the pre-hydrodynamic model and the hydrodynamic simulation. On the other hand, there is meaningful physics associated with the parameter: it is the timescale where the transition occurs. Therefore, the distinction between physical or unphysical (``nuisance'') parameters is context-dependent, and we will not generally make this distinction.
	
	\subsection{Measurements and observables}
	
	Detectors at the Relativistic Heavy Ion Collider and the Large Hadron Collider can measure the momentum distribution of particles, as well as correlations between the particles of different momentum ranges or species. Measurements are grouped by the type of nuclei collided and the center-of-mass energy of the collision: for example, collisions of lead nuclei at center-of-mass energy per nucleon pair of 5.02 TeV (Pb-Pb $\sqrts=5.02$~TeV). The center-of-mass energy $\sqrts$ at the Relativistic Heavy Ion Collider is $\sqrts \leq 0.2$~TeV while most data at the Large Hadron Collider have $\sqrts$ at least 10 times larger ($\sqrts=2.76$~TeV, $\sqrts=5.02$~TeV, $\sqrts=5.44$~TeV). Each collider has multiple detectors and associated experimental collaborations that perform the measurement (ALICE, ATLAS, CMS, PHENIX, STAR, \ldots), and many measurements are performed separately by multiple collaborations. 

	Because of different detector designs and analysis choices, similar measurements are not always trivial to compare. For example, when measuring the number of protons produced on average in heavy-ion collisions, one may choose to exclude or not protons that originate from certain decay processes~\cite{Abelev:2008ab,Adler:2003cb}, leading to substantially different results for observable that could naively be expected to be similar. Another example is momentum cuts: the minimum momentum threshold for particles to be included in a measurement typically varies across detectors, which can lead to somewhat different results for measurements of the ``same'' observable by different collaborations.

	As for the output of the multistage models of heavy-ion collisions, it is typically a shower of particles from which any observable can be calculated. The primary challenge is statistics: some observables require high statistics that are not necessarily straightforward to achieve in numerical simulations, and even less in a Bayesian inference, when simulations must be repeated for a large number of values of the model parameters. Another challenge can be to match precisely the definition of the observable from experimental collaboration. 
	This is a particularly delicate step since a mismatch in evaluating observables will lead to incorrect model-to-data comparisons which would likely go unnoticed.

	\subsection{Model-to-data comparison with Bayesian inference}
	
	Given a choice of model and a set of data, the aim of model-to-data comparisons is to determine (i) whether the model can provide a satisfactory description of the data, and (ii) constrain the value of the model parameters that are consistent with the data and their uncertainty. Related goals include pinpointing which measurements exhibit tension with the model, comparing side-by-side different models (Section~\ref{sec:model_selection}), and identifying observables that are particularly sensitive to specific parameters (experimental design, discussed in Section~\ref{sec:expt_design}).

Bayesian inference provides a framework to extract probabilistic constraints on the model parameters. The key step is the choice of a likelihood function, which quantifies the probability that the observables computed from the model are consistent with the measurement for a given set of model parameter values. In general, a Gaussian likelihood function is used, which in its simplest form (neglecting correlations between observables) is given by
\begin{equation}
	\mathcal{L}(\{ \yexpj \} |\mathbf{p}) =\frac{\exp\left(-\frac{1}{2} \sum\limits_j \frac{\left[ \ythj(\mathbf{p}) - \yexpj \right]^2} {\sigmathj(\mathbf{p})^2 + \sigmaexpj^2}\right)}{\sqrt{\prod\limits_j \left[ (2\pi) \left (\sigmathj(\mathbf{p})^2 + \sigmaexpj^2\right) \right] }} \; .
	\label{eq:likelihood_nocorr}
\end{equation}
where 
\begin{itemize}
	\item $j$ is an index denoting a given data point in the data set;
	\item $\yexpj$ and $\sigmaexpj$ are, respectively, the experimental value and the uncertainty of the data point $j$ (we postpone the discussion of statistical and systematic uncertainties); 
	\item $\ythj(\mathbf{p})$ and $\sigmathj(\mathbf{p})$ are the value and the uncertainty of the model's prediction for the observable corresponding to data point $j$, evaluated when the model's parameters are set to $\mathbf{p}$; the uncertainty  $\sigmathj(\mathbf{p})$ can be numerical, for example from discretization error in a differential equation solvers, or can be statistical, from averaging an insufficient number of stochatic simulations.
\end{itemize} 
Note that Eq.~\ref{eq:likelihood_nocorr} does not group data points in any specific way, a consequence of neglecting correlations between observables and their uncertainties. We discuss this approximation in greater detail in Section~\ref{sec:likelihood}. Also note that the likelihood Eq.~\ref{eq:likelihood_nocorr} is Gaussian in the \emph{observables}, and \emph{not} in the parameters; deviations of the likelihood from a normal distribution can be a reflection of the non-linear dependence of the observables on the model parameters.

The likelihood $\mathcal{L}(\{ \yexpj \} |\mathbf{p})$ (Eq.~\ref{eq:likelihood_nocorr}) is used to define the posterior $\mathcal{P}(\mathbf{p} | (\{ \yexp \})$, which is the probability that the model parameters $\mathbf{p}$ are consistent with the measured data set:
\begin{equation}
	\mathcal{P}(\mathbf{p} | (\{ \yexpj \}) = \frac{\mathcal{L}(\{ \yexp \} |\mathbf{p}) \textrm{Prior}(\mathbf{p})}{\int d\mathbf{p} \mathcal{L}(\{ \yexp \} |\mathbf{p}) \textrm{Prior}(\mathbf{p}) }
	\label{eq:posterior}
\end{equation}
where the prior $\textrm{Prior}(\mathbf{p})$ represents pre-existing constraints on the model parameters~$\mathbf{p}$. These constraints can be theoretical (positivity or other bounds), for example. We discuss priors in more detail in Section~\ref{sec:priors}. The posterior $\mathcal{P}(\mathbf{p} | (\{ \yexp \})$ represents probabilistic constraints on the model parameters. It is a probability distribution with the same dimensions as the number of model parameters $\mathbf{p}$. 

Per se, the theory of model-to-data comparison with Bayesian inference is straightforward, with Eqs~\ref{eq:likelihood_nocorr} and \ref{eq:posterior} summarizing a simple but usable version of the approach. Many of the challenges are numerical, since the observables $\ythj(\mathbf{p})$ are rarely known analytically, and may be slow to evaluate numerically, especially when the number of model parameters is large.

The constraints provided by $\mathcal{P}(\mathbf{p} | (\{ \yexpj \})$ can be summarized by projecting this high-dimensional distribution in lower-dimensional spaces. For example, a probability distribution for the hydrodynamic initial time $\tauhydro$ can be obtained by marginalizing  $\mathcal{P}(\mathbf{p} | (\{ \yexp \})$ over all parameters except $\tauhydro$:
\begin{equation}
	\mathcal{P}(\tauhydro | (\{ \yexpj \}) = \int \prod_{p_\alpha \neq \tauhydro } \!\!\! d p_\alpha \;\; \mathcal{P}(\mathbf{p} | (\{ \yexpj \})
	\label{eq:marginalized_posterior}
\end{equation}
Marginalizing over all but two parameters provides another projection of the posterior that can help visualize correlations between parameters. The single-parameter and two-parameter marginalized distributions are often displayed as triangular matrix plots that are seen in many Bayesian publications. 

A key strength of Bayesian inference is that the model-to-data comparison should be reproducible exactly. This does require all modelling decisions to be explained clearly, ideally accompanied by open-access computer codes (see Section~\ref{sec:data_management}).
In Section~\ref{sec:bayes}, we discuss seven recent Bayesian inferences that constrained the shear and the bulk viscosity as well as other parameters. Most of these analyses obtained different constraints on the viscosity. As we will see, there are multiple reasons for this:
\begin{itemize}
	\item The model is almost always different, even when very similar modelling choices appear to be made;
	\item Different data sets are used;
	\item The priors are different, including but not limited to different ranges of parameters used to compare with data. Differences in the prior can be particularly subtle when trying to constrain a function like the temperature-dependent viscosities, where the prior can be difficult to visualize and compare, as discussed in Section~\ref{sec:bayes_soft_sector};
	\item Correlations between uncertainties (the covariance matrix discussed in Section~\ref{sec:likelihood}) are treated differently;
	\item Numerical uncertainties are present in the results, either in the model calculations themselves, in the emulation (see below), or possibly in the Monte Carlo sampling of the high-dimensional posterior distributions.
\end{itemize}

We comment briefly on the last point, which is directly related to the concept of emulation. 
	
	\subsection{Emulation}
	
	In heavy-ion collisions, the output of the model $\{\ythj(\mathbf{p})\}$ --- the observables of the model --- is only known numerically. Evaluating the posterior requires repeated evaluation of the likelihood $\mathcal{L}(\{ \yexpj \} |\mathbf{p})$ and consequently of the model outputs $\{\ythj(\mathbf{p})\}$ at different values of the parameters $\mathbf{p}$. It is nevertheless understood that $\{\ythj(\mathbf{p})\}$ is a relatively smooth function of the parameters. This smoothness assumption can be used in different ways, but in heavy-ion collisions, it has always been used to replace model observables $\{\ythj(\mathbf{p})\}$ in Eq.~\ref{eq:likelihood_nocorr} by a faster ``emulator'' or a ``surrogate model''. 

\begin{figure}[htbp]
		\begin{center}
			\includegraphics[width=0.7\textwidth]{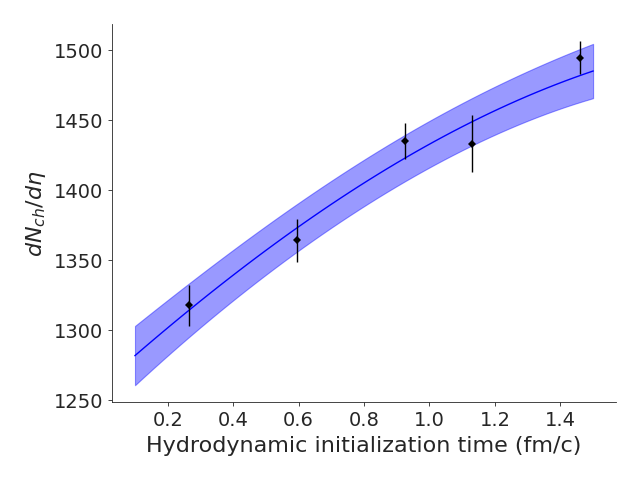}
			\caption{Charged hadron multiplicity $dN_{ch}/d\eta$ as a function of the hydrodynamic initialization time $\tauhydro$, with all the other model parameters kept fixed to the maximum a posteriori parameters for the Grad viscous correction from Ref.~\cite{JETSCAPE:2020mzn}. The black points are the model's output for five different values of $\tauhydro$, with statistical uncertainties originating from averaging over a finite number of collisions. The blue line and band show a probabilistic emulator (a Gaussian process emulator, see Section~\ref{sec:gaussian_processes}) fitted to the model calculations, with the line and the band respectively corresponding to the mean and the 1-$\sigma$ uncertainty of the emulator.}
			\label{fig:emulation_example_tauhydro}
		\end{center}
	\end{figure}

	For example, if we want to study the hydrodynamic initialization parameter $\tauhydro$, we need to know the range of the parameter that we want to study, and have at least some idea of how smooth the model dependence is with respect to $\tauhydro$. If a reasonable range for $\tauhydro$ is $[0.1,1.5]$~fermi and we know from tests that the model's output doesn't change too much when $\tauhydro$ is changed by $0.3$~fermi, then we expect to be able to build an excellent estimator for the model's dependence on $\tauhydro$ with perhaps five evaluations of the model at spaced-out values of  $\tauhydro$. An example is shown in Figure~\ref{fig:emulation_example_tauhydro}: with a small number of model calculations, it is possible to construct an emulator that captures the model dependence on the parameter by interpolating between a small number of calculations while also accounting for the statistical uncertainty of the stochastic model.
	
	This picture becomes much more complicated with a large number of model parameters. 
	Instead of trying to quantify the model's smoothness in 10 or 20 dimensions, we evaluate the model's observables at a few hundred different sets of model parameters (see Section~\ref{sec:sampling_parameter_space}). Using these calculations, we build an emulator that interpolates the model's prediction across the parameter space and estimates the uncertainty in this interpolation. An emulator that accurately captures its own interpolation uncertainty eliminates the need for a ``perfect'' emulator whose uncertainty is negligible compared to the other uncertainties (statistical, numerical, experimental) in the problem. The emulator uncertainty is simply included alongside the other uncertainties. The consequence is that the constraints on the model parameters --- posterior --- have non-negligible uncertainties from the emulator, which can be one more source of difference between Bayesian inference performed by different groups.
	
	While the primary purpose of emulation is often in the service of Bayesian inference, there is value in the emulators themselves. Emulators can help better understand the observable's dependence on the parameters. Emulators can further be used to perform sensitivity analysis of the model, to help understand which observable can best constrain certain sets of parameters, and also understand correlations between parameters. We discuss this in greater detail in the next section. Finally, emulators have an important open-science component: they can summarize the parameter dependence of a complex physical model in a simple, compact object, making it possible to share with the community the results of thousands of hours of calculations. Online tools to visualize the dependence of observable on the model parameters have already been made available, for example Ref.~\cite{jetscape_widget} (see also Refs~\cite{chun_widget,dan_widget}).

	\section{Emulation and applications to heavy-ion collisions}
	
	\label{sec:emulation}
	
	Most of this section covers the practicalities of building an emulator for a model. The last subsection discusses physics questions that can be studied more easily when a model emulator is available.

	\subsection{Model emulation - overview}
	
	Our physical model of heavy-ion collisions has several parameters that characterize unknown or uncertain quantities. These parameters can be the hydrodynamic initialization time $\tauhydro$, the temperature-dependent shear viscosity over entropy density ratio $(\eta/s)(T)$, the temperature-dependent transverse momentum transferred between hard and soft partons $\hat{q}(T)$, etc. When all model parameters are set to a certain value $\mathbf{p}_l$, the model is ``run'' and eventually observables can be evaluated. These observables can be the identified hadron multiplicity, the momentum anisotropy of jets or hadrons, etc. We use the index $j$ to label the different observables, and the index $l$ to identify the set of parameter $\mathbf{p}_l$: $\{\ythj(\mathbf{p}_l)\}$. Here, we consider different bins in momentum or different centrality as different observables. For example, the pion, kaon and proton multiplicities in 10 different centrality bins correspond to 30 observables in our definition. Using this accounting scheme, hundreds of observables have been measured in heavy-ion collisions by different collaborations. Brute-force approaches to emulation could require hundreds of emulators to map the model parameters $\mathbf{p} _l $ to each model output $\{\ythj(\mathbf{p}_l)\}$. There are at least two reasons to avoid this. First, the numerical cost of evaluating emulators scales linearly with the number of emulators, and having hundreds of emulators will slow down the calculation of the posterior, thus reducing the benefits of replacing the model's output with emulators. Second, there are generally strong correlations between most of these observables, and these correlations can be used to improve the emulation by helping differentiate statistical fluctuations from the true dependence of the observables on the model parameters. To reduce the number of emulators, the community has generally been using principal component analysis as a dimensionality reduction tool, which we discuss in Section~\ref{sec:dimensionality_reduction}. 
	
	A separate question is sampling the parameter space $\mathbf{p}_l$. We need to evaluate the model outputs at enough parameter points so that the emulator can adequately capture the observable's dependence on the parameter. The goal is not only to predict the model outputs but also to estimate the interpolation uncertainty in this prediction.
	To sample the parameter space, most groups have used different versions of Latin hypercubes, although adaptive sampling methods have been increasingly explored and used. We review this topic in Section~\ref{sec:sampling_parameter_space}.
	
	Note that we will not distinguish between emulator error and emulator uncertainty in this work. The former is the difference between the emulator's prediction and the real model's output. The latter is the emulator's own estimate of the emulator error --- its interpolation uncertainty in parameter space. We will discuss how to validate the emulator, and we will assume that any emulator used in the discussion has been carefully validated to confirm that it correctly captures its own error; this validation is crucial.
	
	In practice, in heavy-ion collisions, it is almost never the case that emulator uncertainties are negligible compared to the experimental uncertainties.
	There are multiple reasons for this, one being the relatively high accuracy of measurements. Another reason is the complexity of multistage simulations of heavy-ion collisions: when more computing time is available, one generally tries to improve the model and relax modelling assumptions, rather than compute the emulator to high accuracy with a simpler model. At any rate, the presence of at least \emph{some} emulator uncertainty in most Bayesian analyses does imply that the experimental data are not always used to their full extent.

	\subsection{Dimensionality reduction of the observables}
	
	\label{sec:dimensionality_reduction}
	
	Suppose that we are interested in three observables, the pion, kaon and proton multiplicities, in 10 different centrality bins. A brute-force approach would require 30 emulators for these observables, and would ignore the strong correlations between the observables. For example, increasing the total energy of the initial conditions of hydrodynamics will increase all observables similarly. There should be a way to build fewer emulators and reconstruct the relevant information from it.

	Dimensionality reduction can in principle be done by hand. Using existing knowledge of the observables, or by studying the correlation between the observables, one could establish a list of observables with limited redundancy, and restrict model-to-data comparisons to these observables. This is not an outlandish approach, given that such a sorting of the data does not need to be repeated with every new analysis. 
	On the other hand, most observables are merely correlated, not fully redundant, and there would be an inevitable loss of information in such a selection of observables.

	In general, automated method of dimensionality reduction are used, identifying combinations of model outputs (observables) that capture most of the relevant information about the observable's dependence on the model parameter. In the heavy-ion community, this has traditionally been achieved with principal component analysis.
	Detailed descriptions of principal component analyses as applied to heavy-ion physics can be found in Refs~\cite{Bernhard:2018hnz,Moreland:2019szz,coleman2019topics,Heffernan:2023kpm}. We summarize here the main ingredients.
	
	Take $\ythj(\mathbf{p}_l)$, the $j$'th observable that we are interested in, evaluated at the $l$'th set of parameters.
	Over the range of parameters probed, each observable will take a range of values. One can compute the observable's mean and standard deviation over the parameter space. By subtracting the mean and normalizing by the standard deviation, one can first define ``standardized'' observables. This is not yet part of principal component analysis, but rather a pre-processing step. Principal component analysis is then performed to find linear combinations of standardized observables which are called principal components. The principal components are then ordered by how much of the variation over the parameter space each of them explains. In general, a small fraction of the principal components explains most of the variation of the model outputs over the parameter space, and the remaining ones contain so little information that they are effectively indifferentiable from pure noise. One can then emulate only a handful of the dominant principal components and replace the rest with noise terms (see Refs~\cite{Bernhard:2018hnz,Moreland:2019szz} for a detailed explanation).

	\subsection{Sampling of the parameter space and emulation uncertainties}
	
	\label{sec:sampling_parameter_space}
	
	How many times must the model be run with different parameter sets such that we can calculate with confidence all observables at any point in the pre-determined parameter space? Over what range of parameters should the emulator be accurate? Importantly, how should the value of the parameter points be selected?
	
	In the heavy-ion community, these questions have generally been answered as follows:
	\begin{itemize}
		\item Sample the parameter space $10$ to $40$ times the number of parameters; with $20$ parameters, that would mean $200$ to $800$ parameter samples.
		\item Select the value of the parameters based on a Latin hypercube algorithm that spreads out the parameter points across the parameter space.
		\item Limit the parameter space to the hypercube defined by the step-function priors.
	\end{itemize}
	
	Based on emulator validation  performed by different groups (the process of verifying if the emulator faithfully captures the observables' dependence on the model parameters, Section~\ref{sec:emulator_validation}), the procedure described above works relatively well, and yields emulator uncertainty from a few percent to perhaps 20\%, depending on the observable and the analysis. 
	
	There are practical reasons to go beyond the procedure described above. The first reason is the range of the parameter space, which we discuss greater details in Section~\ref{sec:prior_and_sampling}. Second, many Latin hypercube algorithms have the property that they are defined for a fixed number of parameter samples: to distribute the points evenly in parameter space, one must know ahead of time how many points are sampled. This is not necessarily practical. There are versions of Latin hypercubes that have been used in recent studies~\cite{Heffernan:2022swr,Heffernan:2023gye} that at least attempt to order the samples such that e.g., the first 10\% of the samples are distributed over the entire parameter space, and not just in a corner. There also exist versions of Latin hypercubes, such as sliced Latin hypercube, that do not have the requirement that the number of parameter samples is fixed beforehand (see e.g., Ref.~\cite{qian2012sliced}).
	
	The simplicity of Latin hypercubes is that they require no information about the model and the measurements. However, using information about either the model or measurements --- adaptive sampling discussed below --- should allow for more efficient sampling than Latin hypercubes, in the sense that one can have a smaller emulator uncertainty with the same number of samples.

	\subsubsection{Prior and sampling of the parameter space}
	
	\label{sec:prior_and_sampling}
	
	Priors define the probability that a parameter takes a certain value \emph{before} comparison with new measurements. Parameter values that have a finite probability of being consistent with data should be included in the analysis. In previous discussions of sampling (Figure~\ref{fig:emulation_example_tauhydro} for example), we did not factor in the concept of probability, since we were assuming ``step-function priors'' that have a 100\% probability within some range $p_{\alpha, min}<p_\alpha<p_{\alpha, max}$, and 0\% probability otherwise.
	These step-function priors have been used in almost every Bayesian study of heavy-ion collisions, with the notable recent exception of Refs.~\cite{Heffernan:2022swr,Heffernan:2023gye,Heffernan:2023kpm,Heffernan:2023utr}.
	
	We discuss priors in greater details in Section~\ref{sec:priors}, but for the purpose of this section, we highlight the role of priors to guide where parameters should be sampled in the parameter space. It is common in this case to sample the parameter space according to the probability distribution given by the priors. 
	
	\subsubsection{Balancing interpolation uncertainty and statistical uncertainty}
	
	Heavy-ion collision models are stochastic. Averaging over more nucleus collisions (called ``events'') reduces the statistical uncertainty of the observables.  Take Figure~\ref{fig:emulation_example_tauhydro} as an example: it is clear that the black points representing the model calculations have finite statistical uncertainties.
	Two factors that control the uncertainty in the emulator (blue band) are the statistical uncertainty of the model calculations used to train the emulator, and the number of parameter points at which the model calculations are performed. There is an unavoidable trade-off between the two: given a fixed computational budget, increasing the number of parameter samples means fewer events per parameter point, which implies larger statistical uncertainty at each parameter point. The simple example shown in Figure~\ref{fig:emulation_example_tauhydro} has a rather smooth dependence on the model parameters. In this case, fewer parameter samples and more events per parameter sample would be the better choice. Increasing the number of parameter points, even at equal statistical uncertainties, would not significantly improve Figure~\ref{fig:emulation_example_tauhydro}'s emulator.
	
	Nevertheless, one must be very careful with this trade-off, especially with a larger number of parameters. It is challenging to know how smooth the parameter dependence is for a hundred different observables as a function of 10 or 20 parameters. Evaluating the model's prediction at a large number of parameter points is essential if we want the emulator to be able to identify non-linear and non-monotonic behaviour of the model as a function of parameters. From this point of view, statistical uncertainty is better behaved, and there is less of a risk of missing important features of the model because of statistical uncertainty than because of interpolation uncertainty.
	
		\begin{figure}[htbp]
		\begin{center}
			\includegraphics[width=0.5\textwidth]{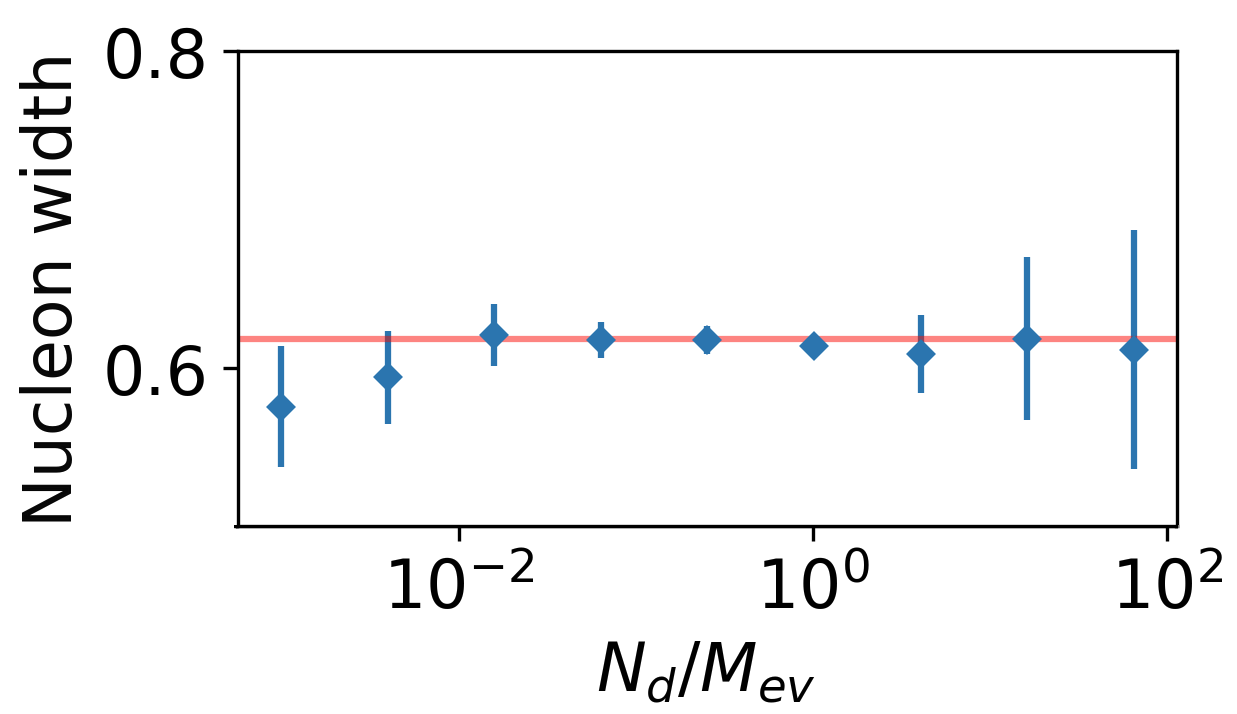}
			\caption{Constraints on one model parameter (the nucleon width from the Trento model~\cite{Moreland:2014oya}) as a function of the ratio between the number of parameter samples $N_d$ and the number of collisions simulated per parameter samples $M_{\textrm{ev}}$, with $N_d M_{\textrm{ev}}$ kept fixed. The horizontal line indicates the real value of the parameter as used in the closure test (Section~\ref{sec:closure_tests}). The points are the mode of the single-parameter marginalized posterior distribution (Eq.~\ref{eq:marginalized_posterior}), while the uncertainty bar of each point is the standard deviation of the distribution.  The total number of simulated collisions $N_d M_{\textrm{ev}}$ represents the available computational budget. A small ratio $N_d/M_{\textrm{ev}}$ implies that few parameters are sampled but many collisions are simulated for each parameter, leading to a small statistical uncertainty and a large interpolation uncertainty. A large $N_d/M_{\textrm{ev}}$ leads to the opposite, a large statistical and a small interpolation uncertainty. All choices of $N_d/M_{\textrm{ev}}$ are consistent with the true value of the parameter, but an optimum is found when a compromise is found between the statistical and emulator uncertainty, around $N_d/M_{\textrm{ev}}\approx 0.1$--$1$ in this test. Figure adapted from Ref.~\cite{Weiss:2023yoj}, see text and reference for details.}
			\label{fig:stat_vs_interp_uncert}
		\end{center}
	\end{figure}
	
	The optimal balance of interpolation and statistical uncertainty depends on the model, especially how non-linear it is, which is itself related to the range of parameters that are studied. If the observables are studied in a very small range of parameters, they will often have a near-linear dependence on the parameters, while probing the model in large ranges of parameter values can expose complex non-linear behaviour. As for the statistical uncertainty, it varies widely from one observable to the next. This can create a competition between observables: those with large statistical uncertainty may benefit from fewer parameter samples and more simulated collisions, and vice versa. Furthermore, there is the experimental uncertainty of each measurement, which varies considerably across observables and represents the true external benchmark to compare the emulator uncertainty with: large emulator uncertainties are not an issue if an observable's measurement is poor and has a large \emph{experimental} uncertainty.
	
	Given all these factors, it is difficult to provide a simple rule to guide the selection of the number of parameter samples and simulated collisions. A starting point is the existing literature: by now, multiple emulators have been prepared for multistage models of heavy-ion collisions, and emulator validation (Section~\ref{sec:emulator_validation}) has been performed on these emulators. Previous studies indicate a satisfactory ability of emulators to capture the parameter dependence of a large number of observables measured in heavy-ion collisions. These previous studies generally used between 500 and 1000 parameter samples distributed with a Latin hypercube algorithm, and 1000 to 10,000 simulated collisions per parameter sample. These choices are unlikely to be optimal but at least provide usable emulators. Nevertheless, one must be careful to make decisions based on previous studies if (i) different observables are used, (ii) new parameters are introduced or a different parameter range is studied, and (iii) the model is modified. 
	
	Aside from existing studies and from trial and error, simplified models can be used to better understand the trade-off between statistical and interpolation uncertainty. Ref.~\cite{Weiss:2023yoj} used a simple model to show that the balance between interpolation and statistical uncertainties could change drastically the degree of constraints on the model parameters, at equal numerical cost. In that study, which focused on anisotropy observables, the optimal balance between uncertainties was found to be when the number of parameter samples was slightly smaller or equal to the number of simulated events, as illustrated in Figure~\ref{fig:stat_vs_interp_uncert}.

	\subsubsection{Static and adaptive sampling}
	\label{sec:adaptive_sampling}
	
	The previous section discussed the challenges of optimizing the emulator uncertainty given a fixed computational budget. A key element of optimizing the emulator uncertainty is selecting the position of the parameter samples in the parameter space. It is useful to divide sampling algorithms into three broad categories: (i) algorithms like Latin hypercubes that do not use any information about the model or the measurements to select the parameter samples; (ii) algorithms that use knowledge of the model but not of the measurements; (iii) algorithms that use the measurements to select the parameter samples.
	
	The first category of algorithms, which include Latin hypercubes, can be referred to as ``static''. The benefits of different Latin hypercube algorithms have not been explored in much detailed yet in the field, with most studies following the lead of Refs~\cite{Bernhard:2018hnz,Bernhard:2019bmu}. Recently, Refs~\cite{Heffernan:2023kpm,Heffernan:2022swr} explored low-discrepancy sequences like Sobol's and the Maximum Projection (MaxPro) Latin hypercube strategy~\cite{joseph2015maximum}, using the latter in a Bayesian analysis~\cite{Heffernan:2023gye,Heffernan:2023utr}. The field could benefit from moving towards these approaches, since they can provide increased convergence of the emulator at little to no cost.
	
	In the rest of this section, we discuss non-static approaches, which can be referred to as ``dynamical''' or ``adaptive''. Evidently, measurements are always involved in the selection of the range of parameters (the priors): for example, there is no point evaluating the model for parameter sets that produce a hundred times more particles than the measurements. This is \emph{not} what we mean by adaptive sampling: we mean automated algorithms that place the point given a certain prior for the parameters.
	
	The general concept of adaptive sampling is that a larger number of parameter samples should be located where:
	\begin{itemize}
	\item the observables have a non-linear dependence on the parameters;
	\item the observables are close to the experimental data;
	\end{itemize}
	The first point is primarily about the efficiency of learning the observable dependence on the parameters.
	Where the model is nearly linear, the parameter samples can be few and far between. More samples are required when observables have a non-linear or even non-monotonic dependence on the parameters. We can think of this as ``model-guided'' adaptive sampling.
	
	As for guiding the selection using measurements, this seems an evident choice: the emulator only needs to be precise for values of the parameters where the model's observables are consistent with the data. On the other hand, to know which parameters can be compatible with the data, one needs a minimum level of emulator accuracy for a wide range of model parameters. This can be thought of as ``measurement-guided'' adaptive sampling.
	
	Various algorithms have been developed to perform adaptive sampling, based on either or both of the principles above~\cite{seo2000gaussian,santner2003design,gramacy2020surrogates,chen2022adaptive,song2023ace}. At least one application to heavy-ion collisions has been made, in Ref.~\cite{Liyanage:2023nds}. 
	
	One benefit of static sampling is the homogeneity of the calculations used to train the emulator. They can all be calculated and organized simultaneously, which is not a trivial benefit when supercomputers are used and millions of files need to be organized. Typically, all parameter samples have similar statistical uncertainties, which makes it possible to use certain simplified emulation techniques. These benefits are not theoretical but practical, and it is one of the reasons that almost all emulators used in heavy-ion collisions up to now have used static sampling. 
	
	Adaptive sampling requires multiple separate rounds of calculations. Parameters could, in principle, be sampled one by one, but for practical reasons, they are generally sampled in batches. Static sampling can be used for the first step, and a small number of parameter samples are drawn. Then, an emulator is built using this initial sample, and a second batch of parameter samples is drawn based on any criterion (strength of observable dependence on the parameters, or level of agreement with measurements). This process is iterated in batches until the computational budget is used or a certain level of emulator accuracy is reached. As discussed in the previous section, the compromise between interpolation and statistical uncertainty must be decided at each step. Ref.~\cite{Liyanage:2023nds} provides a practical demonstration of this approach.
	
	Note that static sampling and model-guided adaptive sampling share one important difference with measurement-guided adaptive sampling. In the former case, the aim is generally to achieve an even emulator uncertainty across the parameter space. In measurement-based adaptive sampling, the emulator uncertainty is expected to be much smaller in regions of the parameter space where the model observables agree with the data, with the emulator uncertainty being potentially much larger elsewhere.

	\subsection{Emulation with Gaussian process regressors}
	
	\label{sec:gaussian_processes}
	
	We discussed in Section~\ref{sec:dimensionality_reduction} that we rarely need a single emulator per observable. Rather, we can first perform dimensionality reduction, such as principal component analysis, on the ensemble of observables and then use a small number of emulators for the dominant principal components. In Section~\ref{sec:sampling_parameter_space}, we surveyed different techniques to select values of model parameters at which the model is evaluated. Up to now, we have kept the concept of emulators relatively general. For a review of emulation in heavy-ion physics, such generality is not necessary, since almost all emulators used in heavy-ion physics are Gaussian process emulators.
	
	Extensive descriptions of Gaussian process regressors are available in other publications~\cite{williams2006gaussian}, including heavy-ion physics publications~\cite{Bernhard:2018hnz,Everett:2021ruv}. In what follows, we only briefly review general aspects.
	
	Assume that there is a single observable $\ythzero(\mathbf{p})$ to emulate.\footnote{Multivalued Gaussian processes do exist and are the subject of active research~\cite{liu2018remarks}.} It is not important for this discussion whether this is an actual observable, a standardized observable (see Section~\ref{sec:dimensionality_reduction}) or a principal component.
	This observable $\ythzero(\mathbf{p})$ has been sampled a certain number of times at different values of the parameters $\mathbf{p}$ (Section~\ref{sec:sampling_parameter_space}). The aim is to construct a continuous function of $\mathbf{p}$ that approximates well the observable and accurately quantifies its statistical and interpolation uncertainty. The Gaussian process is a probability distribution rather than a function.\footnote{Strictly speaking, ``probability distribution'' is reserved for countable number of degrees of freedom, while ``random process'' or ``stochastic process'' is used for objects such functions that have an infinite number of degrees of freedom.} This probability distribution is assumed to be a multidimensional Gaussian (``multivariate normal distribution''), such that it can be specified from two moments only: its mean and its covariance. The only information available to constrain the mean and the covariance are (i) the sampled parameters that we denote $\{\mathbf{p}_l\}$, (ii) corresponding calculated values of the observables $\{\ythzero(\mathbf{p}_l)\}$ and (iii) the statistical uncertainty of each sample $\{\Delta\ythzero(\mathbf{p}_l)\}$. The different $\{\ythzero(\mathbf{p}_l)\}$ evaluated at the parameter sets $\{\mathbf{p}_l\}$ have some degree of correlation, which is related to how smooth the function is. We write the covariance of the observable at two different parameter points $\mathbf{p}_l$ and $\mathbf{p}_k$ as
	\begin{equation}
		\textrm{covariance}(\ythzero(\mathbf{p}_l),\ythzero(\mathbf{p}_m))=k(\mathbf{p}_l,\mathbf{p}_m)+\sigma_n^2 \delta_{lm}
		\label{eq:covariance_noise}
	\end{equation}
	where $k(\mathbf{p}_l,\mathbf{p}_m)$ is the \emph{covariance kernel} and $\sigma_n$ is a noise term that is meant to account for the statistical uncertainty in the calculations. One way of understanding Gaussian process regressors is that, once the covariance kernel $k(\mathbf{p}_l,\mathbf{p}_m)$ is selected, one can write the mean prediction of the Gaussian process at an arbitrary parameter $\mathbf{p}^*$ as the sum of the covariance between every parameter sample~\cite{williams2006gaussian}
	\begin{equation}
		\mu(\mathbf{p}^*)=\sum_{l=1}^{N_s} \alpha_l k(\mathbf{p}_l,\mathbf{p}^*)
	\end{equation}
	with $\alpha_l$ given by a linear superposition of the $\ythzero(\mathbf{p}_l)$. 
	
	The choice of covariance function is an important assumption, and using a covariance function that is not a good match for the model can lead to considerable problems with the emulator, such as overfitting or underfitting, or associating statistical fluctuations with physical features of the model. Conversely, a well-chosen covariance function can improve considerably the performance of the emulator.
	
	A common choice for the covariance kernel $k(\mathbf{p}_l,\mathbf{p}_m)$, that has been used in many applications to heavy-ion collisions, is the generic \emph{squared exponential kernel}
	\begin{equation}
		k(\mathbf{p}_l,\mathbf{p}_m)=\sigma^2_f \exp\left(-\frac{1}{2}\sum_a \frac{\left|p_{l,a}-p_{m,a}\right|^2}{\ell_a^2}\right)
		\label{eq:squared_exponential}
	\end{equation}
	which assumes a strong correlation between the model outputs when the values of the parameters are close compared to an external length scales $\ell_a$, with $p_{l,a}$ being the ``a'''th parameter of the parameter vector $\mathbf{p}_l$. In theory, if the parameter dependence of the model is already known well,  one has a good idea of the length scales $\ell_a$ over which the model outputs change appreciably. In practice, these length scales $\ell_a$ are determined by numerical optimization to the model outputs $\{\ythzero(\mathbf{p}_l)\}$;  $\sigma_f$ and $\sigma_n$ are determined by numerical optimization as well. This is called ``hyperparameter determination''. More information can be found in Refs~\cite{williams2006gaussian,Everett:2021ruv}.
	
	Note that Eq.~\ref{eq:covariance_noise} assumes that the noise term is independent of the parameter points: that is, the statistical uncertainty is assumed to be the same across the parameter space. This is called homoscedastic noise. This may not be the case in general, in particular if adaptive sampling is used, leading to heteroscedastic noise. Ref.~\cite{Liyanage:2023nds} provides an example of handling this in applications to heavy-ion collisions.
	
	Gaussian process regressors can take more complex forms than the version presented above. So-called multitask Gaussian process emulators, which predict multiple observables simultaneously, have been explored in some works~\cite{Soeder:2023vdn} and have considerable potential to improve the emulation of correlated observables. Another recent study in Ref.~\cite{Li:2023ize} used a combination of Gaussian processes, rather than a single one, as emulator; this ``multi-index'' approach can take advantage of the known multistage structure of heavy-ion models, such that different Gaussian process can attempt to capture the physics of specific stages of the collision.
	
	There is an important reason for the field's focus on Gaussian process emulators: it is necessary to use an emulation method that provides both an accurate prediction of the model's outputs, and an accurate determination of the interpolation uncertainty. That is, it is essential for the emulator to quantify its own limitations. Unless emulation uncertainties can be reduced significantly below experimental uncertainties --- a very challenging task at the moment --- Bayesian inference is dependent on a reliable quantification of the emulator uncertainty, which are included in the final uncertainty on the model parameters. Because of this constraint, there has been limited exploration of other emulation techniques, although it will be important to investigate further.

	\subsection{Emulator validation}
	
	\label{sec:emulator_validation}
	
	Replacing a model's outputs with emulators requires high confidence that the emulators accurately capture the behaviour of the original physical model.
		
	Emulator validation is inseparable from the discussion of sampling of the parameter space in Section~\ref{sec:sampling_parameter_space}. Unless quantitative external information about the model is available, all our knowledge comes from evaluating its outputs (observables) at different sets of parameters. 
	Evidently, emulator validation must also be viewed in terms of sampling the parameter space with a finite computational budget.
	
	A typical emulator validation approach is to divide the model calculations at different parameter samples into two groups\footnote{Note that ``training'', ``validation'' and ``testing'' have a somewhat different meaning in machine learning~\cite{khan2014alpaydin}.}: (i) a ``training set'' used to prepare the emulator, and (ii) a separate ``validation set''. The emulator is trained using the first set, and its ability to describe observables is quantified with the validation set. Metrics used to quantify the agreement can be as straightforward as the mean-squared error between the emulator prediction and the calculations from the validation set. Other metrics used in the machine learning community have not been used regularly yet in the heavy-ion community, but can be found in the literature (see e.g., Refs.~\cite{chung2021beyond, chung2021uncertainty,PSAROS2023111902} and references therein).
	
	\begin{table}
		\begin{center}
			\begin{tabular}{@{} cccccc @{}}
				\hline
				Training set & 1,2,3,4 & 1,2,3,5 & 1,2,4,5 & 1,3,4,5 & 2,3,4,5 \\ 
				\hline
				Validation set & 5 & 4 & 3 & 2 & 1 \\ 
				\hline
			\end{tabular}
		\end{center}
		\begin{center}
			\begin{tabular}{@{} ccccccccccc @{}}
				\hline
				Training set & 1,2,3 & 1,2,4 & 1,2,5 & 1,3,4 & 1,,3,5 & 1,4,5 & 2,3,4 & 2,3,5 & 2,4,5 & 3,4,5 \\ 
				\hline
				Validation set & 4,5 & 3,5    & 3,4    & 2,5    & 2,4 & 2,3 & 1,5 & 1,4 & 1,3 & 1,2 \\ 
				\hline
			\end{tabular}
		\end{center}
		\caption{Illustration of leave-p-out cross-validation with $p=1$ (top) and $p=2$ (bottom) and 5 total samples.}
		\label{table:drop_k_validation}
	\end{table}
	
	When using static parameter sampling (e.g., Latin hypercube), separate Latin hypercubes can be used for the training and validation sets, with a larger number of samples in the training set. Validation can also be performed without clear separation between the sets: the emulator is trained using all but $k$ parameter samples, and the rest are used for validation, iterating over all possibilities. The 5-parameter-sample case is illustrated in Table~\ref{table:drop_k_validation} for $k=1$ and $k=2$, although in general, the total number of parameters is evidently much larger than $k$. This is referred to as ``leave-p-out cross-validation''. A discussion of this type of validation applied to heavy-ion collisions can be found in Ref.~\cite{Bernhard:2018hnz}. Although the use of separate sets of training and validation parameters has been relatively common in the field, the ``leave-p-out cross-validation'' is likely to provide better validation in general.
	
	Emulator validation with adaptive sampling methods is more subtle because adaptive sampling itself is a recursive attempt to estimate the emulator uncertainty and error. As discussed in Section~\ref{sec:sampling_parameter_space}, static sampling and model-guided adaptive sampling generally aim for uniform emulator uncertainty across the parameter space. Measurement-guided adaptive sampling allows for broad variation in the emulator uncertainty, with the emulator only needing to be highly accurate in parameter ranges where the model's observables are close to the experimental measurements. Emulator validation must take this into account.
		
	\subsection{New developments in emulation}
	
	\label{sec:emulation_new_devs}
	
	In application to heavy-ion physics, our general goal for emulators is to capture as much physics as possible using the smallest possible computational budget. Capturing more physics can mean including additional effects in the models of heavy-ion collisions, or describing as many systems as possible (different nuclei collision and center-of-mass energies). The previous sections have discussed some tools that can help minimize the computational time necessary to prepare emulators from a model. We briefly survey a selection of other recent developments in what follows. 
	
	\subsubsection{Transfer learning}
	
	When multiple models are similar but not identical, it is possible to prepare emulators for each one at a reduced cost, by exploiting their similarity. In that case, the parameter dependence of the model's outputs can be complicated, but the parameter dependence of one model's output \emph{relative} to the other is expected to be modest. Thus, we can build multiple model emulators by dividing the problem in two: first, train an emulator for one of the models (referred to as the \emph{source}), and second, build emulators for the difference between the first model and the other ones (referred to as the \emph{targets}).
	
	\begin{table}[htb]
		\begin{center}
			\begin{tabular}{@{} |c|c|c| @{}}
				\hline
				Paper & Source emulator & Target emulator \\ 
				\hline
				Ref.~\cite{Liyanage:2022byj}  & Au-Au $\sqrts=200$~GeV & Pb-Pb $\sqrts=2760$~GeV \\ 
				\hline
				Ref.~\cite{Liyanage:2022byj}  & Grad viscous correction & Chapman-Enskog \& P.T.B viscous correction \\ 
				\hline
				Ref.~\cite{Heffernan:2023utr}  & Grad viscous correction  & Chapman-Enskog viscous correction  \\ 
				\hline
			\end{tabular}
		\end{center}
		\caption{Examples of applications of transfer learning in modelling of heavy-ion collisions.
		}
		\label{table:transfer_learning}
	\end{table}
	
	Applications to heavy-ion physics of this approach were introduced in Ref.~\cite{Liyanage:2022byj}, using transfer in two different cases: transfer from one collision system to another, and transfer across different models of transition from hydrodynamics to post-hydrodynamics (Grad, Chapman-Enskog, P.T.B.; see Ref.~\cite{McNelis:2019auj} and references therein). Building on Ref.~\cite{Liyanage:2022byj}, Refs~\cite{Heffernan:2023utr,Heffernan:2023gye} used transfer learning in a full-fledged Bayesian inference. These applications are summarized in Table~\ref{table:transfer_learning}. In both papers, the authors found that transfer learning produced emulators at a fraction of the cost of training full emulators from scratch. 
	
	\subsubsection{Multifidelity with non-ordered models}

	\begin{figure}[tb]
		\centering
		\begin{tikzpicture}[node distance=2.5cm]
			\node (prehydro) [rectangle, draw, fill=gray!10, text width=4.5cm, align=center] {Pre-hydrodynamics};
			\node (optical) [right of=prehydro, node distance=5cm, text width=4cm, align=center] {Optical Glauber model};
			\node (eventbyevent) [right of=optical, node distance=5cm, text width=4cm, align=center] {Event-by-event model with initial transverse flow};
			\node (simple) [above of=eventbyevent, node distance=1.5cm, rectangle, draw, text width=4.5cm, align=center] {High-fidelity model};
			\node (simple) [above of=optical, node distance=1.5cm, rectangle, draw, text width=4.5cm, align=center] {Low-fidelity model};
			\node (hydro) [below of=prehydro, rectangle, draw, fill=gray!25, text width=4.5cm, align=center] {Hydrodynamics};
			\node (ideal) [right of=hydro, node distance=5cm, text width=4cm, align=center] {Ideal hydro};
			\node (viscous) [right of=ideal, node distance=5cm, text width=4cm, align=center] {Viscous hydro with temperature-dependent shear and bulk};
			\node (posthydro) [below of=hydro, rectangle, draw, fill=gray!40, text width=4.5cm, align=center] {Post-hydrodynamics};
			\node (hadronic) [right of=posthydro, node distance=5cm, text width=4cm, align=center] {Hadronic decays};
			\node (transport) [right of=hadronic, node distance=5cm, text width=4cm, align=center] {Hadronic transport};
			
			\draw [->] (prehydro) -- (hydro);
			\draw [->] (hydro) -- (posthydro);
			\draw [->] (optical) -- (ideal);
			\draw [->] (ideal) -- (hadronic);
			\draw [->] (eventbyevent) -- (viscous);
			\draw [->] (viscous) -- (transport);
			
		\end{tikzpicture}
		\caption{Comparison of a simpler (low-fidelity) and a more complete (high-fidelity) multistage models of heavy-ion collisions.}
		\label{fig:hic_model_old_new}
	\end{figure}
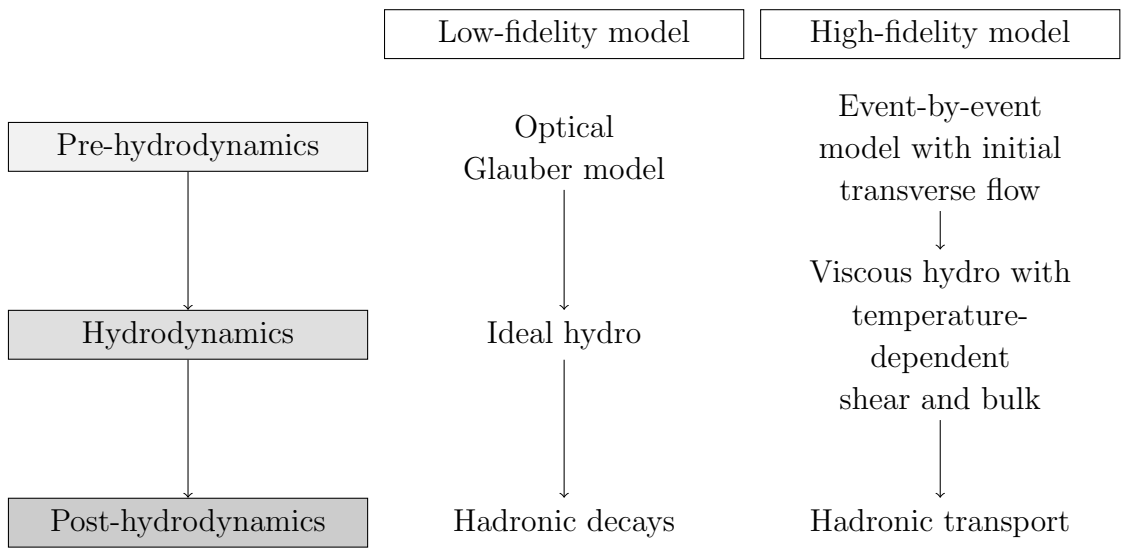
	
	Multifidelity emulation can be illustrated in heavy-ion physics with the example shown in Figure~\ref{fig:hic_model_old_new}, which compares a modern multistage model of heavy-ion collisions (``high-fidelity model'') with a simple model used in the earlier days of the field (``low-fidelity model'').
	We discussed above that transfer learning can be used to transfer information from one emulator (the source) to another (the target) when the models are similar.
	The general idea of multifidelity modelling is the same, except that the ``source'' model is a version of the full model that is much less expensive to calculate, and hence assumed to be simpler as well (lower fidelity).
	The idea is again that the source model captures a lot of the parameter dependence of the target model, and that preparing an emulator that captures the difference between the two models is easier. Whether multifidelity emulation is cost-efficient depends on the relative cost of the full model and the simpler one, and the degree of similarity between the models.
	
	A more advanced approach combines multiple simpler models, each providing information about the final model at a reduced cost. If the simpler models can be ordered strictly from least accurate (low fidelity) to most accurate (high fidelity), then one is doing \emph{ordered} multifidelity emulation~\cite{kennedy2000predicting}. If the relation between the models' accuracy isn't clear (for example, ``optical Glauber + viscous hydrodynamics + hadronic decay'' compared to ``event-by-event initial conditions + ideal hydro + hadronic decay''), then one must do unordered multifidelity, which was the topic of Ref.~\cite{ji2021graphical}.

	\subsubsection{Multifidelity with accuracy parameter extrapolation}
	
	An evident application of multifidelity emulation is for models that have numerical accuracy parameters, such as time steps, spatial grid size, \ldots  In this case, one does not need multiple models of variable speed and fidelity --- as in Figure~\ref{fig:hic_model_old_new} --- to benefit from multifidelity emulation. One can use a single model, and take calculations with larger numerical error as a``low-fidelity'' emulator that captures a significant fraction of the full model's parameter dependence.

	It might appear that there is no need for a special approach to do multifidelity emulation such as this, since the time step or the spatial grid step might simply be added as additional model parameters. This simple approach may work to an extent, but fundamentally, it is important to differentiate between numerical and model parameters. Emulators attempt to \emph{interpolate}, in the parameter space, the dependence of observables on model parameters, assuming that the observables have a somewhat smooth parameter dependence. There is, however, no implicit assumption that the model reaches a plateau or converges to a specific value. For numerical accuracy parameters, we do assume that the observables should reach an asymptote. In fact, we would ideally want to \emph{extrapolate} the observables to this asymptote even if we cannot calculate the observable in this idealized limit of negligible numerical error. In the language of Section~\ref{sec:gaussian_processes}, this means that the covariance kernel that should be used for numerical parameters is different than for other parameters.
	An approach that can achieve this goal, applied to heavy-ion physics, was presented in Ref.~\cite{Ji:2022xzo}.

	\subsubsection{Nonparametric approach for functional model parameters}
	
	A particular challenge of Bayesian inference in heavy-ion collision is the need to extract functions rather than scalar parameters: the temperature dependence of the shear and bulk viscosity, the temperature and momentum dependence of the hard parton momentum broadening $\hat{q}(T)$, the temperature and baryon chemical potential dependence of the nuclear equation of state, etc.
	
	From a physical point of view, a partial solution is to reduce the variation of these parameters with proper dimensional scaling. For example, the shear viscosity $\eta$ is understood to depend considerably on the temperature of the plasma, but the ratio of the shear viscosity to the entropy density, $\eta/s$, has a smaller dependence, and is even constant in certain theories. Similarly, the temperature dependence of the parton momentum broadening $\hat{q}$ should be reduced considerably by taking the ratio to the entropy density or the cube of the temperature, $\hat{q}/s$ or $\hat{q}/T^3$.
	
	Even after these rescaling, these quantities still have a non-negligible temperature dependence. To extract this temperature dependence from measurements, a parametrization must be selected. For bulk viscosity, for example, all recent studies assumed a parametrization of the form of a generalized Cauchy distribution
	\begin{equation}
		\frac{\zeta}{s}(T) = \frac{(\zeta/s )_{\max}}{1+ \left( \frac{T-T_\zeta}{w_{\zeta} \left[1 + \lambda_{\zeta} \textrm{sign} \left(T{-}T_\zeta\right) \right]}\right)^2}
		\label{eq:zeta_over_s_jetscape}
	\end{equation}
	with some groups assuming $ \lambda_{\zeta}=0$ (symmetric peak of $(\zeta/s)(T)$) and others a non-zero  $\lambda_{\zeta}$ (asymmetric peak). As we will discuss in Section~\ref{sec:bayes_soft_sector}, these choices are modelling assumptions that can have considerable effects on the resulting extraction of the viscosity.

	A different method is to use a non-parametric approach, like a Gaussian process emulator, to constrain the functional dependence of the parameter. One of the benefits is that it does not necessarily introduce long-range correlations in the parametrization. Take the specific bulk viscosity $(\zeta/s)(T)$ of Eq.~\ref{eq:zeta_over_s_jetscape} with  $\lambda_{\zeta}=0$, such that the parametrization is a symmetric peak around a certain temperature $T_\zeta$.  If the value of $(\zeta/s)(T)$ at a temperature below the peak is strongly constrained, then it automatically provides strong constraints at some temperature above the peak. If there is strong evidence that $(\zeta/s)(T)$ is symmetric, this is a useful strong modelling assumption; if not, it leads to spurious correlations for the values of $(\zeta/s)(T)$ at different temperatures. A Gaussian process emulator can avoid these correlations by requiring a certain smoothness between neighbouring values of $(\zeta/s)(T)$ without introducing long-range correlations. This degree of correlation can be controlled by changing the correlation function or its hyperparameters (for example, Eq.~\ref{eq:squared_exponential}). The effectiveness of this approach was shown in Ref.~\cite{Xie:2022ght} for constraints on the temperature dependence of the parton momentum broadening $\hat{q}$. Another benefit of the approach is that it is straightforward to extend to multivariate functions.

	\subsection{Uses of model emulators}
	
	\label{sec:emulation_applications}
	
	While we are primarily discussing emulators in the context of Bayesian inference, we briefly mention here applications of model emulators that are somewhat independent of inference.
	
	\subsubsection{Sensitivity analysis and correlations}
	
	An important factor in our ability to constrain a model parameter is how strongly the model's observables depend on it, and whether different parameters lead to similar effects in the observables (correlations and degeneracies between parameters). In the example shown in Figure~\ref{fig:emulation_example_tauhydro} , the charged hadron multiplicity changes by approximately $20\%$ over a reasonable range of values for the hydrodynamic initialization time $\tauhydro$. Whether this $20\%$ is large or small depends on the size of the experimental uncertainties: if the charged hadron multiplicity can be measured with sub-percent accuracy, then one would expect a strong constraint on the hydrodynamic initialization time, while a $20\%$ experimental uncertainty would result in a relatively weak constraint. Strong conclusions about the parameter dependence of different observables ultimately require discussing experimental data. On the other hand, one can go far with limited experimental information. For example, sub-percent experimental uncertainties are rare in heavy-ion physics once all sources of uncertainty are accounted for, and any parameter that requires this level of accuracy is unlikely to be constrained strongly after a full-fledged Bayesian inference.
	
	Moreover,  it is possible to understand correlations between model parameters in great detail by studying the model somewhat independently from measurements. Figure~\ref{fig:emulation_example_tauhydro} might suggest the hydrodynamic initialization time is straightforward to constrain with sufficiently small experimental uncertainties on the multiplicity, but multiple parameters have similar effects on this observable.
	
	Ref.~\cite{Liyanage:2022byj} provides a recent example of sensitivity analysis, which used Sobol' indices~\cite{sobol2001global} to study the dependence of model observables on the different parameters. Of note is the use of grouped Sobol' indices to quantify the impact on the observables of e.g., all parameters related to bulk viscosity, rather than each parameter one by one. Other sensitivity analyses include Refs~\cite{Sangaline:2015isa,JETSCAPE:2020mzn,Parkkila:2021yha,Everett:2021ruv,Heffernan:2023utr,Heffernan:2023kpm}.

	\subsubsection{Visualization of parameter dependence of observables}
	
	Exploring the parameter dependence of a model (Section~\ref{sec:sampling_parameter_space}) can require days or weeks of calculations on modern computer clusters. An emulator using these calculations can summarize these results into a simple object that can be shared with the community. An example is the release of the emulators from Refs~\cite{JETSCAPE:2020mzn,JETSCAPE:2020shq} in the form of an interactive website where users can change the value of the parameter and observe the effect on observables~\cite{jetscape_widget}. This provides additional transparency to the publications, since others can readily access the emulator (see Section~\ref{sec:data_management} for a broader discussion). Moreover, multistage models of heavy-ion collisions are complex, and it is often not straightforward to associate the effect of a physical process with an observable, or to understand the compensating impact of parameters on observables. Direct online access to the emulators can help readers test and understand the observables' parameter dependence independently. Other publications have released similar online tools for other models~\cite{chun_widget,dan_widget}.

	\section{Bayesian inference: general concepts}
	
	\label{sec:bayes}
	
	\begin{figure}[tb]
		\begin{center}
			\includegraphics[width=0.7\textwidth]{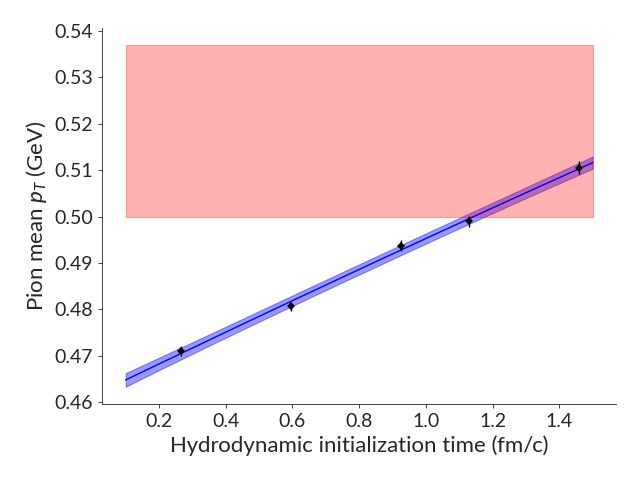}
			\caption{Pion mean transverse momentum ($p_T$) as a function of the hydrodynamic initialization time $\tauhydro$, with all the other model parameters fixed as described in Figure~\ref{fig:emulation_example_tauhydro}. The black points are the model's output~\cite{JETSCAPE:2020mzn} for five different values of $\tauhydro$, with statistical uncertainties. The blue line and band show a probabilistic emulator (a Gaussian process emulator, see Section~\ref{sec:gaussian_processes}) fitted to the values, with the line corresponding to the mean of the emulator and the band to the one-sigma uncertainty. The red band is the experimental uncertainty (Pb-Pb $\sqrt{s_{NN}}=2.76$~TeV, 0-5\% centrality~\cite{ALICE:2013mez}).}
			\label{fig:emulation_example_meanpT_tauhydro}
		\end{center}
	\end{figure}
	
	Experimental collaborations invest considerable resources in quantifying measurements' statistical and systematic uncertainties. These uncertainties propagate unto the model parameters of interest, such as the viscosity of the plasma, the hydrodynamic initialization time, etc. This is illustrated in Figure~\ref{fig:emulation_example_meanpT_tauhydro}, which shows one observable --- pion mean transverse momentum ($p_T$) --- as a function of the hydrodynamic initialization time. The model calculations are shown as black points, and they can be seen to have small statistical uncertainties, as well as a clear trend as a function of hydrodynamic initialization time. The emulator shown by the blue line and band captures this clear trend and has very little interpolation uncertainty --- the blue band primarily captures the statistical uncertainty. The measurement~\cite{ALICE:2013mez} and its uncertainty partly overlap with the model calculations. We see that the largest hydrodynamic initialization time $\tauhydro=1.5$~fm has the best agreement with the pion mean $p_T$, while $\tauhydro=1.2$~fm is a less likely value although certainly not an excluded one. In fact, only rather small values of $\tauhydro$ could be excluded at ``two sigmas''.\footnote{Statements like this do assume that the uncertainty is Gaussian, which is not necessarily correct, especially for systematic uncertainties.} Based solely on this measurement, the probabilistic constraint on $\tauhydro$ would thus be relatively broad.
	
	Performing model-to-data comparison with Bayesian inference allows for methodical uncertainty quantification. We can use Figure~\ref{fig:emulation_example_meanpT_tauhydro} to make some observations:
	\begin{itemize}
	\item Because there are multiple model parameters, we cannot determine the constraints on the hydrodynamic initialization time $\tauhydro$ solely from Figure~\ref{fig:emulation_example_meanpT_tauhydro}; by modifying other parameters, the model's prediction for the pion mean $p_T$ will change at a fixed $\tauhydro$. Constraints are obtained by varying all parameters simultaneously;
	\item While the emulator uncertainty in Figure~\ref{fig:emulation_example_meanpT_tauhydro} is small, this is not always true, and constraints on the parameters will often have some dependence on the emulator's statistical and interpolation uncertainty, that is, uncertainties other than the experimental ones;
	\item No distinction is made between statistical and systematic uncertainties in Figure~\ref{fig:emulation_example_meanpT_tauhydro}, but this distinction can be important, especially when multiple observables are used and they are correlated.
	\end{itemize}

	In this section, we present Bayesian inference at a level necessary to understand Bayesian inference in heavy-ion collisions, starting with a discussion of priors and likelihoods and moving to more specialized topics.
	
	\subsection{Bayesian inference}
	
	The probabilistic constraints on the model parameters $\mathbf{p}$ from the measurements $\{ \yexp \}$ are given by the posterior $\mathcal{P}(\mathbf{p} | (\{ \yexp \})$, Eq.~\ref{eq:posterior}, which we repeat here:
	\begin{equation}
		\mathcal{P}(\mathbf{p} | (\{ \yexp \}) = \frac{\mathcal{L}(\{ \yexp \} |\mathbf{p}) \textrm{Prior}(\mathbf{p})}{\int d\mathbf{p} \mathcal{L}(\{ \yexp \} |\mathbf{p}) \textrm{Prior}(\mathbf{p}) }
		\label{eq:posterior_2}
	\end{equation}
	where $\textrm{Prior}(\mathbf{p})$ is the parameter prior and $\mathcal{L}(\{ \yexpj \} |\mathbf{p})$ is the likelihood function, which we discuss in turn.
	
	\subsubsection{Priors}
	
	\label{sec:priors}
	
	The model parameters entering into numerical simulations of heavy-ion collisions are not wholly unconstrained. Priors refer to any of these pre-existing constraints. Priors can have multiple origins, including (i) theoretical guidance, (ii) previous comparisons with measurements, and (iii) model self-consistency arguments.
	
	As an example, we can take the shear viscosity over entropy density at zero baryon chemical potential, $(\eta/s)(T)$. In this case, the prior is not a scalar but a continuous ensemble of functions of temperature.
	The following considerations can be used to establish a prior:
	\begin{itemize}
		\item At a minimum, we know that $(\eta/s)(T)$ is positive definite, and the prior can exclude any negatively-valued functions;
		\item $(\eta/s)(T)$ should be a continuous function, given that the thermodynamics indicates a crossover rather than a first- or second-order phase transition (at zero baryon chemical potential)~\cite{Borsanyi:2013bia, Bazavov:2014pvz};
		\item Large values of $(\eta/s)(T)$ can likely be excluded by model self-consistency arguments: in this limit, the effect of viscosity on the hydrodynamic evolution can become unjustifiably large, and one can even reach a point where the hydrodynamic evolution is acausal~\cite{Bemfica:2020xym,Plumberg:2021bme,Chiu:2021muk,ExTrEMe:2023nhy};
		\item Calculations exist for $(\eta/s)(T)$ in different limits, including at low temperatures ($\approx 100$~MeV~\cite{Lu:2011df,Romatschke:2014gna,Pratt:2016elw, Rose:2017bjz, Rose:2020lfc}) and at very high temperatures~\cite{Arnold:2003zc,Arnold:2006fz,Demir:2008tr,Ghiglieri:2018dib} (perhaps hundreds of GeV), and they suggest that that  $(\eta/s)(T)$ reaches a minimum in the crossover region;
		\item There are indications from the study of other substances that $(\eta/s)(T)$ reaches a minimum in the crossover region~\cite{Csernai:2006zz};
		\item There are different arguments suggesting a minimum value $(\eta/s)(T)$ larger than $0$, perhaps $(\eta/s)(T) \gtrsim 1/4\pi$~\cite{Kovtun:2004de} (see Ref.~\cite{Schaefer:2014awa} for a review).
	\end{itemize}
	Even limiting ourselves to the above information, establishing a prior is challenging. Should one strictly enforce $(\eta/s)(T) > 1/4\pi$ despite the imperfect theoretical foundations of this limit? To what extent should the prior be constrained by calculations of $(\eta/s)(T)$ at low and high temperatures, given that these calculations' exact regime of validity is unclear? Should the prior assume that there is a single minimum of $(\eta/s)(T)$ with the function being strictly monotonic away from this minimum? Note the priors for $(\eta/s)(T)$ is not necessarily independent from the prior on other parameters: whether a certain parametrization of $(\eta/s)(T)$ leads to acausal hydrodynamic evolution will depend on the bulk viscosity and the initial condition parameters, for example.
	
	\begin{figure}[htb]
	\begin{center}
		\includegraphics[width=0.5\textwidth]{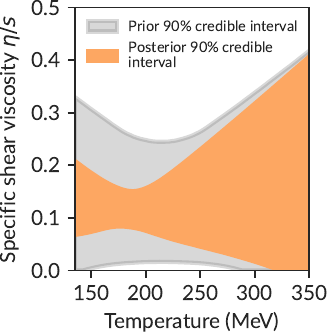}
		\caption{Prior and posterior for the specific shear viscosity (shear viscosity over entropy density) as constrained from RHIC and LHC measurements in Ref.~\cite{JETSCAPE:2020mzn}. The similarity of the prior and posterior at high temperature shows that there is little constraints from data at high temperature; the constraints  --- for example, $\eta/s$ not being larger than $0.4$ --- is largely from the assumed prior. Figure adapted from Ref.~\cite{JETSCAPE:2020mzn}.}
		\label{fig:prior_posterior}
	\end{center}
\end{figure}
	
	Because definitions of priors are rarely clear-cut, priors can be seen as modelling assumptions. It used to be common to neglect the effect of bulk viscosity in simulations of heavy-ion collisions, which is equivalent to defining a null prior for bulk viscosity. For shear viscosity, many would agree with the assumption that $\eta/s(T)$ has a single minimum and has a monotonic behaviour away from this minimum, at least in the temperature range relevant for heavy-ion collisions. For some, enforcing this seemingly reasonable modelling assumption improves the Bayesian inference since it excludes parameterizations of $\eta/s(T)$ that are thought to be unphysical. For others, enforcing this assumption deprives us of verifying if any parametrization that does not satisfy this property can still describe measurements.
	
	Deciding on modelling assumptions provides general guidance but not absolute constraints on the priors. Because of the difficulty in defining priors,
	the use of step-function priors ($p_{j, min}<p_j<p_{j, max}$) should generally be avoided unless (i) there exists a strict bound on the parameter (such as positivity), or (ii) one wants to enforce a modelling assumption strictly. In general, modelling assumptions are imprecise (for example, ``very large $\eta/s(T)$ should be avoided''), and priors should continuously taper off to zero.

	A simple approach to assess the role played by the prior is to overlay it with the posterior, as shown in Figure~\ref{fig:prior_posterior}. If the prior and posterior are the same for a certain range of parameters, it highlights that the constraints obtained are largely from the prior itself, and not from data. This is not necessarily an issue if the prior is well motivated, but in the opposite case, one may need more data or a re-evaluation of the theoretical basis behind the choice of prior.
	
	Priors determined using previous comparisons with measurements are an important but subtle case. In fact, posteriors from Bayesian inference are almost never used \emph{directly} as priors in heavy-ion physics. An example where a previous \emph{posterior} could be reused as \emph{prior} is if one first compares with a set of data ``A'', and then wants to compare with a second set of data ``B'' (which excludes any data from ``A''). This is not necessarily essential since comparing with data sets ``A'' and ``B'' simultaneously can be more straightforward in practice. However, there can certainly be cases where measurement-informed priors are a good approach.
	
	There are, however, multiple cases where measurement-informed priors constitute a problem. The primary reason is that different groups tend to make different modelling assumptions and thus tend to obtain different posteriors even when comparing with similar data sets (see Section~\ref{sec:bayes_soft_sector}). In this case, comparing a model with data sets ``A'' and ``B'' simultaneously is more accurate than attempting to use another model's posterior as prior.  Doing so also removes the possibility of using the same data set twice.

	\subsubsection{Likelihood}
	
	\label{sec:likelihood}
	
	We noted in Section~\ref{sec:brief_intro} that Gaussian likelihoods are the ones that have been used historically for Bayesian inference in heavy-ion collisions, as in many other fields. The general form of the Gaussian likelihood is
	\begin{equation}
		\mathcal{L}(\{ \yexpj \} |\mathbf{p}) =\frac{\exp\left(-\frac{1}{2} \sum\limits_{j,k=1}^{N_{exp}} \left[ \ythj(\mathbf{p}) - \yexpj \right] \Sigma^{-1}_{j k}(\mathbf{p}) \left[ \ythk(\mathbf{p}) - \yexpk \right] \right) }{\sqrt{(2\pi)^{N_{exp}}  \det[\Sigma(\mathbf{p})]}}
		\label{eq:likelihood}
	\end{equation}
	where 
	\begin{itemize}
		\item $j$ and $k$ are indices denoting a given data point in the data set;
		\item $N_{exp}$ is the total number of data points;
		\item $\yexpj$ is the mean of the experimental data point $j$; 
		\item $\ythj(\mathbf{p})$ is the value of the model's prediction for the observable corresponding to data point $j$ evaluated when the model's parameters are set to $\mathbf{p}$;  if an emulator is used, then $\ythj(\mathbf{p})$ is replaced by the emulator's output;
		\item $\Sigma$ is the covariance matrix.
	\end{itemize} 
	
	Equation~\ref{eq:likelihood_nocorr} is recovered when the covariance matrix is diagonal, $\Sigma_{j k}=\sigma_j^2 \delta_{j k}$.
	
	We focus our discussion on the covariance matrix $\Sigma$, since the interpretation of the other variables is straightforward.
	
	The covariance matrix is a sum of:
	\begin{itemize}
		\item the statistical uncertainty of the calculations;
		\item the interpolation uncertainty of the emulator;
		\item the statistical uncertainty of the experimental measurements;
		\item the systematic uncertainty of the experimental measurements.
	\end{itemize}
	Some of these uncertainties are correlated, and others are not. When there are no correlations between uncertainties, the covariance matrix takes the form
	\begin{equation}
		\Sigma^{(uncorr)}=\textrm{diag}(\sigma_1^2, \sigma_2^2, \ldots, \sigma_{N_{exp}}^2)
	\end{equation}
	When there are correlations, it is useful to first consider the \emph{correlation} matrix $C$. The diagonal elements of the correlation matrix $C$ are $1$. The off-diagonal elements depend on the degree of correlation between the observables: $1$ if the observables are fully correlated, $-1$ if fully anti-correlated, $0$ if uncorrelated, and intermediate values in other cases. The covariance matrix is related to the correlation matrix by $\Sigma_{j k}=\sigma_j \sigma_k C_{j k}$.
	
	When the model is replaced by Gaussian process emulators, the emulators' full (non-diagonal) covariance matrix is used to account for the interpolation uncertainty, the statistical uncertainty, and any correlations between the observables.
	
	The experimental uncertainties are more challenging, because the covariance matrix is generally not published by experiments. Correlations of uncertainties in centrality or momentum are sometimes specified. Correlations across observables are rarely available. There have been previous attempts to use theoretical information or general guidance to build a model-based experimental covariance matrix for the systematic uncertainty~\cite{Bernhard:2018hnz,JETSCAPE:2021ehl}. The effect of these prescriptions has not been studied much yet. Note that most earlier studies assume that the experimental statistical uncertainties are uncorrelated, which cannot be correct for all observable (for example, the pion multiplicity and the charged hadron multiplicity).

	\subsection{Numerical aspects}
	
	\label{sec:bayes_numerics}
	
	Combining the prior and the likelihood gives the posterior (Eq.~\ref{eq:posterior_2}). Extracting useful information from the posterior requires computing various moments of the distribution, which is high dimensional. 
	Markov chain Monte Carlo (MCMC)~\cite{hastings1970monte} have generally been used for this task.
	The numerical needs of these Markov chain Monte Carlo illustrate the usefulness of model emulators. The number of samples used to evaluate moments of a typical high-dimensional posterior with a Markov chain is in the hundreds of thousands. This would, in theory, require evaluating the model's output at least this number of times (generally much more, depending on the acceptance rate of the Markov chain).
	On the other hand, in heavy-ion collisions, emulators are prepared after sampling the model (Section~\ref{sec:sampling_parameter_space}) at a few hundred to a few thousand parameter sets. 
	Performing a Markov chain Monte Carlo with emulators instead of the model thus typically requires at least two orders of magnitude fewer model calculations.
	
	This approach of building emulators and sampling them with Markov chain Monte Carlo is equivalent to trying to integrate a high-dimensional function by fitting an emulator to the integrand before performing a numerical integration on the fit. While it is not obvious that this is a particularly efficient strategy, one of the approach's strengths is that it factorizes assumptions about the smoothness of the underlying model from the high-dimensional Monte Carlo integration. 
	
	Note that sampling a 10- to 20-dimensional probability distribution is far from obvious, and efforts must be made to ascertain that the draws from the Markov chain Monte Carlo properly sample the posterior distribution. Validation that can be used includes autocorrelation tests; discussions can be found in Refs.~\cite{Everett:2021ruv, Heffernan:2023kpm}.
	
	\subsection{Maximum a posteriori parameters}
	
	The posterior represents the probability that the parameter set $\mathbf{p}$ is consistent with the experimental data for a given model. There is usually a single parameter set where the model achieves the best possible agreement with data: the parameter set where the posterior is maximum, called ``maximum a posteriori parameters'', often shortened informally as ``MAP''. Multiple strictly equal maxima in the posterior are unlikely given the complexity of the model, but the ``uniqueness'' of the maximum a posteriori parameters should not make one overestimate its meaningfulness: it is not uncommon for the posterior to have relatively flat features in the parameter space, meaning that certain parameters are poorly constrained, and a range of parameter values will yield almost indistinguishable agreement with data.
	
	\subsection{Model selection}
	
	\label{sec:model_selection}
	
	As discussed in Section~\ref{sec:brief_intro}, there is not one standard model of heavy-ion collisions. Even within a single approach, there are several possible variations of components of heavy-ion collision models. Model selection in Bayesian inference is the process of comparing two or more models using criteria that account for the model's ability to describe measurements and the number of model parameters, for example, to discourage complicated and poorly predictive models. Discussions of model selection can be found in the literature, in Refs.~\cite{Liddle:2007fy,Liddle:2009xe} for example.
	
	One approach used for model selection is the Bayes factor (see Ref.~\cite{kass1995bayes} for a review). The Bayes factor for model ``$a$'' uses the evidence $E(\mathcal{M}_a)$, which is the likelihood weighted by the prior, integrated over the parameter space (denominator of Eq.~\ref{eq:posterior_2}):
	\begin{equation}
		E(\mathcal{M}_a)=\int d\mathbf{p} \mathcal{L}_{\mathcal{M}_a}(\{ \yexp \} |\mathbf{p}) \textrm{Prior}(\mathbf{p}) \; .
	\end{equation}
	The evidence favours good agreement with measurements but penalizes a lack of predictivity.
	
	Under the simplifying assumption that there is no prior evidence to favour one of the models over the other, the Bayes factor is given by the ratio of the Bayes evidence of two models:
	\begin{equation}
		B(\mathcal{M}_1,\mathcal{M}_2)=\frac{E(\mathcal{M}_1)}{E(\mathcal{M}_2)}
		\label{eq:Bayes_factor}
	\end{equation}
	
	Bayes factors have been studied in a few heavy-ion publications already~\cite{JETSCAPE:2020mzn,JETSCAPE:2020shq,Everett:2021ruv,Heffernan:2023utr}, to compare (i) different models of conversion of hydrodynamic's energy-momentum tensor to particles, (ii) whether certain parameters should be kept fixed for different center-of-mass energies, and (iii) whether measurements favour a temperature-dependent shear viscosity over entropy density ratio, among others.
	
	Other criteria can be used to perform model comparisons, and model selection is evidently an imperfect science. In Refs.~\cite{JETSCAPE:2020mzn,JETSCAPE:2020shq,Everett:2021ruv}, one model was found to be very strongly disfavoured by the Bayes factor compared to other models. A detailed investigation suggested that this model's difficulty in describing the proton-to-pion ratio was the main reason for the model being disfavoured. This is a meaningful observation. On the other hand, the proton-to-pion ratio is an observable that likely has non-trivial modelling uncertainties, in the sense that it would not be surprising if this observable were sensitive to features of the model that have not been thoroughly explored yet. In this sense, model selection must always be interpreted with the proper physics perspective.
	
	Nevertheless, quantitative comparisons between models are important. The number of parameters in models of heavy-ion collisions can vary significantly. Comparing one model's ability to fit data with another cannot make sense without a degree of quantification of the model's predictivity, which is related to its number of parameters.
	
	Moreover, there must be an attempt to quantify the benefit of different parameters. Adding a parameter inevitably leads to a better agreement with data, but this better agreement cannot per se be interpreted as support for adding the parameter. Model selection can help determine if the improvement in describing data brought by a new parameter is sufficient to justify its weakening of the model's predictivity.  Refs~\cite{Heffernan:2023gye,Heffernan:2023utr} provide a recent example, finding that Bayes factors did not favour a temperature-dependent specific shear viscosity over a constant one, because the slight improvement in agreement with the data did not warrant the addition of three extra model parameters (the temperature-dependent viscosity has four parameters, and the constant viscosity has a single one).

	\subsection{Model averaging and mixing}
	
	\label{sec:model_averaging}
	
	Model selection discussed in the previous section does not necessarily imply that specific models are ruled out by comparison with data. It could be that no model is overwhelmingly favoured by model selection. In this case, it can make more sense to combine the constraints obtained with different models.	
	In Bayesian model averaging, the posterior of different models is averaged in proportion to their Bayes factor (Eq.~\ref{eq:Bayes_factor}). Results of model averaging can be found in Refs~\cite{JETSCAPE:2020shq,Everett:2021ruv,Heffernan:2023gye,Heffernan:2023utr}, for example.
	
	A different approach is model mixing, which combines different models based on their strengths to obtain a new model that can be closer to the truth than each individual model.  
	For example, one model might be more accurate in a certain parameter range and another modelling choice better in a different parameter range.
	Another possibility is one model being more accurate for certain centralities, center-of-mass energy or type of colliding nuclei, or other ``experimental control parameter''; in this case, the model mixing is performed using this control parameter, rather than a model parameter.
	In neither of these cases would model averaging be able to account for the differences in the models. A general overview of model mixing and its applications to nuclear physics can be found in Ref.~\cite{Phillips:2020dmw}, and examples and applications to heavy-ion collisions can be found in Refs~\cite{coleman2019topics,Everett:2021ruv,liyanage_thesis}.

	\subsection{Validation with closure tests}

	\label{sec:closure_tests}
	
	Emulator validation was discussed in Section~\ref{sec:emulator_validation}. In theory, sufficient validation has been performed at this previous step to proceed to comparison with experimental data. In practice, emulator validation is always a challenging task, and valuable additional validation can be performed at the Bayesian inference level. 
	
	A closure test means replacing the experimental measurements in Bayesian inference with model calculations. A set of model parameters $\mathbf{p}_{truth}$ is chosen, observables are calculated, a Bayesian inference is performed on the calculated observables, and the posterior is compared with the true model parameters. If the emulator is accurate and quantifies its uncertainty correctly, the true parameters $\mathbf{p}_{truth}$ should be in a high-posterior region. Ideally, the posterior would be sharply peaked around the true value of the parameters, but this will depend on the sensitivity of the observable to the parameter, and on the uncertainties of the emulator and of the model calculations used for the closure test.
	
	Closure tests are a powerful tool to validate the entire analysis before comparison with data, including the sampling of the posterior with Markov chain Monte Carlo. They further provide the basis for experimental design, discussed later in Section~\ref{sec:expt_design}.
	
	The result of closure tests will depend on the choice of $\mathbf{p}_{truth}$. The choice of $\mathbf{p}_{truth}$ follows the same discussion as that of the choice of a validation set in Section~\ref{sec:emulator_validation}.
	
	An interesting question is the quantification of closure tests. Is there a metric to determine how successful closure has been? Again, earlier topics provide partial answers. Comparing the quality of the closure achieved by two different emulators can be done with model selection tools discussed in Section~\ref{sec:model_selection}, as used in Ref.~\cite{Weiss:2023yoj}. We refer the readers to Refs~\cite{Weiss:2023yoj,JETSCAPE:2021ehl,Fan:2023wpv} for other investigations of the quantification of closure tests in the context of heavy-ion collisions.

	\section{Bayesian inference: applications in heavy-ion physics}
	
	\label{sec:bayes_applications}
	
	As stated in the introduction, a certain level of factorization is possible in the study of heavy-ion physics: model-to-data comparison generally does not include all possible measurements at the same time. The study of ``soft physics'' or low-energy particles is primarily concerned with constraining the equation of state, the transport coefficient, and the equilibration of the plasma or other properties of the early stage of the collision. Hard physics focuses on properties that quantify the degree of interaction between high-energy quarks and gluons with lower-energy ones, quantities such as $\hat{q}$. In this section, we first provide a brief overview of applications of Bayesian inference in heavy-ion collisions. We next discuss in detail one example: soft-sector constraints on the shear and bulk viscosity.
	
	\subsection{Overview}
	
	Many current applications of Bayesian inference in heavy-ion collisions have their roots in the early works~\cite{Petersen:2010zt,Bass:2012gy, Novak:2013bqa} by groups at Duke and Michigan State University, many of whom members of the MADAI Collaboration~\cite{madai}. The combination of dimensionality reduction with principal component analysis, emulation with Gaussian process regressors and Markov chain Monte Carlo sampling were all present in Ref.~\cite{Novak:2013bqa}. Note that non-Bayesian approaches were being explored at the same time~\cite{Soltz:2012rk}.
	
	Important early applications of Bayesian inference include Ref.~\cite{Sangaline:2015isa}, which dissected the results of Bayesian inference to better understand the relation between measurements, their uncertainties and constraints on the model parameters --- a form of experimental design (see Section~\ref{sec:expt_design}). Work in Ref.~\cite{Pratt:2015zsa} used measurements from heavy-ion collisions to constrain the equation of state of nuclear matter, finding consistency with first-principle lattice calculations~\cite{Borsanyi:2013bia, Bazavov:2014pvz}.
	
	Building upon these previous works, a series of papers by the Duke group~\cite{Bernhard:2015hxa, Bernhard:2016tnd,Ke:2016jrd,Ke:2018tsh, Moreland:2018gsh, Bernhard:2018hnz, Bernhard:2019bmu} marked a transition to performing Bayesian inference using multistage models of heavy-ion collisions that include most of the elements that are considered ``state-of-the-art'' at the moment: event-by-event simulations, flexible initial conditions, temperature-dependent shear and bulk viscosities, and a hadronic afterburner. The open-source codes written, optimized and published by these authors form the backbone of almost all Bayesian inference performed since. Additional work has been done by collaborations such as JETSCAPE~\cite{JETSCAPE:2020mzn,JETSCAPE:2020shq,JETSCAPE:2021ehl,JETSCAPE:2022cob,coleman2019topics,Everett:2021ruv,Heffernan:2023kpm} and BAND~\cite{Phillips:2020dmw} to provide open-source frameworks that build or expands on these previous efforts.
	
	In Section~\ref{sec:bayes_soft_sector}, we inspect seven different recent analyses that attempted to constrain the shear and bulk viscosity of quark-gluon plasma with soft hadron observables. The aim is to use these analyses as an example of the possibilities and challenges of Bayesian inference, and how seemingly similar models can lead to significantly different results. We emphasize that there are multiple other analyses beyond these seven, including older analyses listed above, as well as more recent ones~\cite{Auvinen:2017fjw,Auvinen:2020mpc,JETSCAPE:2022cob,Yang:2022ixy,Nijs:2022rme,Nijs:2023yab,Vermunt:2023fsr}. There have also been a number of Bayesian inference focused on parton energy loss~\cite{Xu:2017obm,Ke:2018tsh,He:2018gks,Ke:2020clc,JETSCAPE:2021ehl,Xie:2022ght,Wu:2023olb,Fan:2023wpv,Liu:2023rfi,Karmakar:2023ity,Xing:2023ciw}, 
	nuclear structure~\cite{Cheng:2023ucp,Giacalone:2023cet}, 3D initial conditions~\cite{Soeder:2023vdn}, blast-wave fit~\cite{Yang:2018ghi,Yang:2020oig}, the equation of state and transport coefficients at finite baryon chemical potential~\cite{Oliinychenko:2022uvy,Shen:2023awv,Yang:2023apw} and many more.

	\subsection{Bayesian constraints on the viscosity}
	
	\label{sec:bayes_soft_sector}

	\begin{table}[tb]
		\small
		\centering
			\begin{tabular}{@{} |>{\raggedright\arraybackslash}p{1.85cm}|>{\raggedright\arraybackslash}p{2.2cm}|>{\raggedright\arraybackslash}p{1.7cm}|>{\raggedright\arraybackslash}p{1.7cm}|>{\raggedright\arraybackslash}p{2.6cm}|>{\raggedright\arraybackslash}p{2.9cm}| @{}}
				\hline
				\textbf{Reference} & \textbf{Pre-hydro}  & \textbf{Hydro} & \textbf{Cooper-Frye} & \textbf{Data} & \textbf{Covariance} \\ 
				\hline
				Bernhard et al~\cite{Bernhard:2019bmu} & Trento+f.s. & DNMR & P.T.B.; $\sigma$~meson production & Pb-Pb @ 2.76~TeV \& 5.02~TeV & $\Sigma_{emul}$\hspace{1cm}+diag. $\Sigma_{expt}^{stat}$\hspace{1cm}+non-diag. $\Sigma_{expt}^{syst}$\hspace{2cm}+ $\Sigma_{extra}$ \\ 
				\hline
				Moreland et al \cite{Moreland:2018gsh} & Trento w/ subnucleonic d.o.f.+f.s. & DNMR & P.T.B.; $\sigma$~meson production & p-Pb \& Pb-Pb @ 5.02~TeV & $\Sigma_{emul}$\hspace{1cm}+diag. $\Sigma_{expt}^{stat}$\hspace{1cm}+non-diag. $\Sigma_{expt}^{syst}$\hspace{2cm}\\ 
				\hline
				JETSCAPE \cite{JETSCAPE:2020mzn,JETSCAPE:2020shq} & Trento+f.s. & DNMR  & Grad, Chapman-Enskog, P.T.B. & Au-Au @ 0.2~TeV \& Pb-Pb @ 2.76~TeV & $\Sigma_{emul}$\hspace{1cm}+diag. $\Sigma_{expt}^{stat}$\hspace{1cm}+diag. $\Sigma_{expt}^{syst}$\hspace{2cm} \\ 
				\hline
				Nijs et al~\cite{Nijs:2020ors, Nijs:2020roc} & Trento w/ subnucleonic d.o.f. + modified streaming & DNMR & P.T.B.; $\sigma$~meson production & Pb-Pb @ 2.76~TeV; p-Pb \& Pb-Pb @ 5.02~TeV; added differential observables & $\Sigma_{emul}$\hspace{1cm}+diag. $\Sigma_{expt}^{stat}$\hspace{1cm}+non-diag. $\Sigma_{expt}^{syst}$ \\ 
				\hline
				Parkkila et al \cite{Parkkila:2021tqq,Parkkila:2021yha} & Trento+f.s. & DNMR & P.T.B. $\sigma$~meson production & Pb-Pb @ 2.76~TeV \& 5.02~TeV ; added event-plane correlations  & $\Sigma_{emul}$\hspace{1cm}+diag. $\Sigma_{expt}^{stat}$\hspace{1cm}+non-diag. $\Sigma_{expt}^{syst}$\hspace{2cm}+ $\Sigma_{extra}$  \\ 
				\hline
				Liyanage et al \cite{Liyanage:2023nds} & Trento + anisotropic hydro parameters & Viscous anisotropic hydro & P.T.M.A. & Pb-Pb @ 2.76~TeV  & $\Sigma_{emul}$\hspace{1cm}+diag. $\Sigma_{expt}^{stat}$\hspace{1cm}+diag. $\Sigma_{expt}^{syst}$\hspace{2cm}  \\
				\hline 
				Heffernan et al \cite{Heffernan:2023utr,Heffernan:2023gye} & IP-Glasma & DNMR & Grad, Chapman-Enskog & Pb-Pb @ 2.76~TeV; added event-plane correlations & $\Sigma_{emul}$\hspace{1cm}+diag. $\Sigma_{expt}^{stat}$\hspace{1cm}+diag. $\Sigma_{expt}^{syst}$\hspace{2cm}  \\ 
				\hline
			\end{tabular}
			\caption{Summary of relevant features of seven recent Bayesian analyses that constrained the shear and the bulk viscosity. Free-streaming is denoted ``f.s.''. For the covariance column, $\Sigma_{emul}$ is the emulator covariance that contains both the statistical uncertainty and the interpolation uncertainty, while $\Sigma_{expt}^{stat}$ and $\Sigma_{expt}^{syst}$ are to covariance matrix assumed for the experimental statistical and systematic uncertainties, and $\Sigma_{extra}$ is an additional uncertainty added meant to account for the imperfection of the model. See text for more details. Note that Parkkila et al \cite{Parkkila:2021tqq,Parkkila:2021yha} and Heffernan et al \cite{Heffernan:2023utr,Heffernan:2023gye}  added different sets of transverse-momentum-integrated observables --- see the respective publications for more information.}
			\label{table:bayes_viscosity}
		\end{table}

		\begin{table}[tbh]
			\small
			\centering
				\begin{tabular}{@{} |>{\raggedright\arraybackslash}p{3.65cm}|>{\raggedright\arraybackslash}p{11.2cm}|}
					\hline
					Publication & Parametrization of shear and bulk viscosity \\ 
					\hline
					Bernhard et al~\cite{Bernhard:2019bmu} Moreland et al \cite{Moreland:2018gsh} Nijs et al~\cite{Nijs:2020ors, Nijs:2020roc} Parkkila et al \cite{Parkkila:2021tqq,Parkkila:2021yha} &
					
					\begin{equation*}
						\hspace{-2.4cm}\frac{\eta}{s}(T) =
						\left\{ \begin{array}{ll}
							\left(\frac{\eta}{s}\right)_{min}+\left(\frac{\eta}{s}\right)_{slope}(T-T_c)(T/T_c)^{\left(\frac{\eta}{s}\right)_{crv}} & \textrm{if } T>T_c;\\
							(\eta/s)_{HRG} & \textrm{otherwise}
						\end{array}\right.
					\end{equation*}
					with $T_c=154$~MeV.
					\vspace{.2cm}
					\begin{equation*}
						\hspace{-2.4cm}\frac{\zeta}{s}(T) = \frac{(\zeta/s )_{\max}}{1+ \left( \frac{T-T_\zeta}{w_\zeta}\right)^2}
					\end{equation*}

					\\
					\hline
					JETSCAPE \cite{JETSCAPE:2020mzn,JETSCAPE:2020shq}; Liyanage et al \cite{Liyanage:2023nds}; Heffernan et al \cite{Heffernan:2023utr,Heffernan:2023gye}   &
					\begin{equation*}
						\hspace{-2.4cm}\frac{\eta}{s}(T) =
						\left\{ \begin{array}{ll}
							\textrm{max}\left\{ a_{\rm low}\, (T{-}T_{\eta}) , 0 \right\} & \textrm{if } T>T_{\eta};\\
							\textrm{max}\left\{ a_{\rm high}\, (T{-}T_{\eta}) , 0 \right\} & \textrm{otherwise}
						\end{array}\right.
					\end{equation*}
					\vspace{.2cm}
					\begin{equation*}
						\hspace{-2.4cm}\frac{\zeta}{s}(T) = \frac{(\zeta/s )_{\max}}{1+ \left( \frac{T-T_\zeta}{w_{\zeta} \left[1 + \lambda_{\zeta} \textrm{sign} \left(T{-}T_\zeta\right) \right]}\right)^2}
					\end{equation*}
					\\ 
					\hline
				\end{tabular}
				\caption{Parametrization of shear and bulk viscosity over entropy density ratio, for the analyses discussed in this section. All variables except $T$ and $T_c$ are model parameters.}
				\label{table:visc_param}
			\end{table}

			Table~\ref{table:bayes_viscosity} lists seven different analyses that compared multistage models of heavy-ion collisions with low-energy hadronic measurements, and obtained constraints on the temperature dependence of the viscosities. The analyses are (i) Bernhard et al~\cite{Bernhard:2019bmu}; (ii) Moreland et al \cite{Moreland:2018gsh}; (iii) JETSCAPE \cite{JETSCAPE:2020mzn,JETSCAPE:2020shq}; (iv) Nijs et al~\cite{Nijs:2020ors, Nijs:2020roc}; (v) Parkkila et al \cite{Parkkila:2021tqq,Parkkila:2021yha};  (vi) Liyanage et al \cite{Liyanage:2023nds}; and (vii) Heffernan et al \cite{Heffernan:2023utr,Heffernan:2023gye}. 
			While these studies will inevitably be superseded by newer analyses in the future, as models are improved and numerical capabilities increase, they represent a good case study of Bayesian inference applied to heavy-ion physics. The author of this review was involved with two of the analyses: JETSCAPE \cite{JETSCAPE:2020mzn,JETSCAPE:2020shq} and Heffernan et al~\cite{Heffernan:2023utr,Heffernan:2023gye}.
			
			Before proceeding with the comparison, we highlight that the analysis of Liyanage et al \cite{Liyanage:2023nds} has a major difference with the others: it used anisotropic relativistic viscous hydrodynamics~\cite{McNelis:2021zji,McNelis:2018jho} rather than DNMR relativistic viscous hydrodynamics~\cite{Denicol:2012cn}, the M\"uller-Israel-Stewart-type transient viscous hydrodynamics approach used in most comparisons with heavy-ion data. 
			
			In what follows, we discuss in turn the differences between the analyses.\footnote{A discussion and comparison of many of the seven analyses can also be found in Section 2.7.2 of Ref.~\cite{ALICE:2022wpn}.}
						
			\paragraph{Pre-hydrodynamics}
			Six of the seven analyses used Trento initial conditions, although two different versions of Trento were used: both Nijs et al and Moreland et al studied small collision systems (p-Pb) and used a version of Trento with subnucleonic degrees of freedom~\cite{Moreland:2018gsh}, while Bernhard et al, JETSCAPE, Parkkila et al and Liyanage et al did not have subnucleonic fluctuations~\cite{Moreland:2014oya}. One analysis (Heffernan et al) used IP-Glasma initial conditions~\cite{Schenke:2012wb,Schenke:2012fw}.
			
			Trento is generally understood to provide an initialization for the energy or entropy density at very early times (see Ref.~\cite{Nijs:2023yab} for a recent discussion). Because anisotropic hydrodynamics is meant to be applicable at these early times, Trento is used directly to initialize hydrodynamics in Liyanage et al. Note that Liyanage et al does have other initial condition parameters that are set independently from Trento.
			
			In Bernhard et al, Moreland et al, JETSCAPE and Parkkila et al, the Trento initial conditions were combined with a free-streaming phase\footnote{The description of the free-streaming process can be found in Refs~\cite{Moreland:2018gsh,Bernhard:2019bmu} based on Refs~\cite{Liu:2015nwa,Broniowski:2008qk}.} to provide an initialization for the entire energy-momentum tensor of hydrodynamics at a later hydrodynamic initialization time $\tauhydro$. Nijs et al~\cite{Nijs:2020ors, Nijs:2020roc} used a procedure similar to free-streaming but with a variable propagation velocity.
			
			Finally, Heffernan et al \cite{Heffernan:2023utr,Heffernan:2023gye} used IP-Glasma initial conditions to initialize DNMR hydrodynamics at different times.
			
			Note that even after selecting a pre-hydrodynamics model, some freedom remains. One degree of freedom is the pressure: the equation of state of the system enters in the definition of the pressure, but the pre-equilibrium model does not necessarily have the same equation of state as the hydrodynamics (pre-equilibrium models are often conformal, while the equation of state of quark-gluon plasma is only conformal at very high temperatures). The choice of equation of state to define the pressure impacts the bulk pressure used to initialize hydrodynamics~\cite{NunesdaSilva:2020bfs}. As for the shear tensor part of the energy-momentum tensor, the one extracted from a free-streaming pre-equilibrium model is often large, which can contribute to acausal problems in the hydrodynamics evolution at early time~\cite{Plumberg:2021bme,Chiu:2021muk}. While most analyses did use the full energy-momentum tensor of the free-streaming model to initialize hydrodynamics, other options are being explored (see e.g., Ref.~\cite{Nijs:2023yab}). Pre-equilibrium models such as K\o{}MP\o{}ST are meant to improve the matching between free-streaming and hydrodynamics~\cite{Kurkela:2018wud,Kurkela:2018vqr}, although they have not been used yet in Bayesian inference and they do not appear to fully resolve causality issues encountered at early times~\cite{Plumberg:2021bme} or other challenges~\cite{NunesdaSilva:2020bfs}.

			\paragraph{Hydrodynamics, equation of state and transport coefficients}
			
			We already mentioned the choice of hydrodynamics equations: all analyses used DNMR relativistic viscous hydrodynamics~\cite{Denicol:2012cn} except Liyanage et al who used viscous anisotropic hydrodynamics~\cite{McNelis:2021zji,McNelis:2018jho}. The results of the hydrodynamic equations depend on the equation of state and the transport coefficients. 
			
			The equation of state has strong constraints from lattice QCD~\cite{Borsanyi:2013bia, Bazavov:2014pvz}, although energy-momentum conservation at the transition from hydrodynamics to hadronic degrees of freedom requires a forced matching to the hadron resonance gas equation of state~\cite{Anderlik:1998et,Huovinen:2009yb,Moreland:2015dvc,Auvinen:2020mpc}. Most groups now try to match exactly the hadronic content of the hadronic transport to the equation of state; this helps with conservation of energy and momentum across the transition from hydrodynamics to hadronic transport, although uncertainties remain (see Refs.~\cite{Moreland:2015dvc,Auvinen:2020mpc} and also Ref.~\cite{SalinasSanMartin:2022inb} for example). One source of confusion is the physics of the hadron resonance gas. As discussed in detail in Ref.~\cite{JETSCAPE:2020mzn}, there is at least one hadron --- the $\sigma$ meson --- that should not be treated in the same way as the others and that should not be included when computing the thermodynamics of the hadron resonance gas~\cite{Broniowski:2015oha}. As indicated in Table~\ref{table:bayes_viscosity}, four out of the seven analyses did include the $\sigma$ meson in the thermodynamics, modifying the equation of state, and as we discuss below, increasing the number of pions produced by the model. As a result, the equation of state used in the different analyses did vary somewhat.
			
			\begin{figure}[tb]
				\begin{center}
					\includegraphics[width=0.7\textwidth]{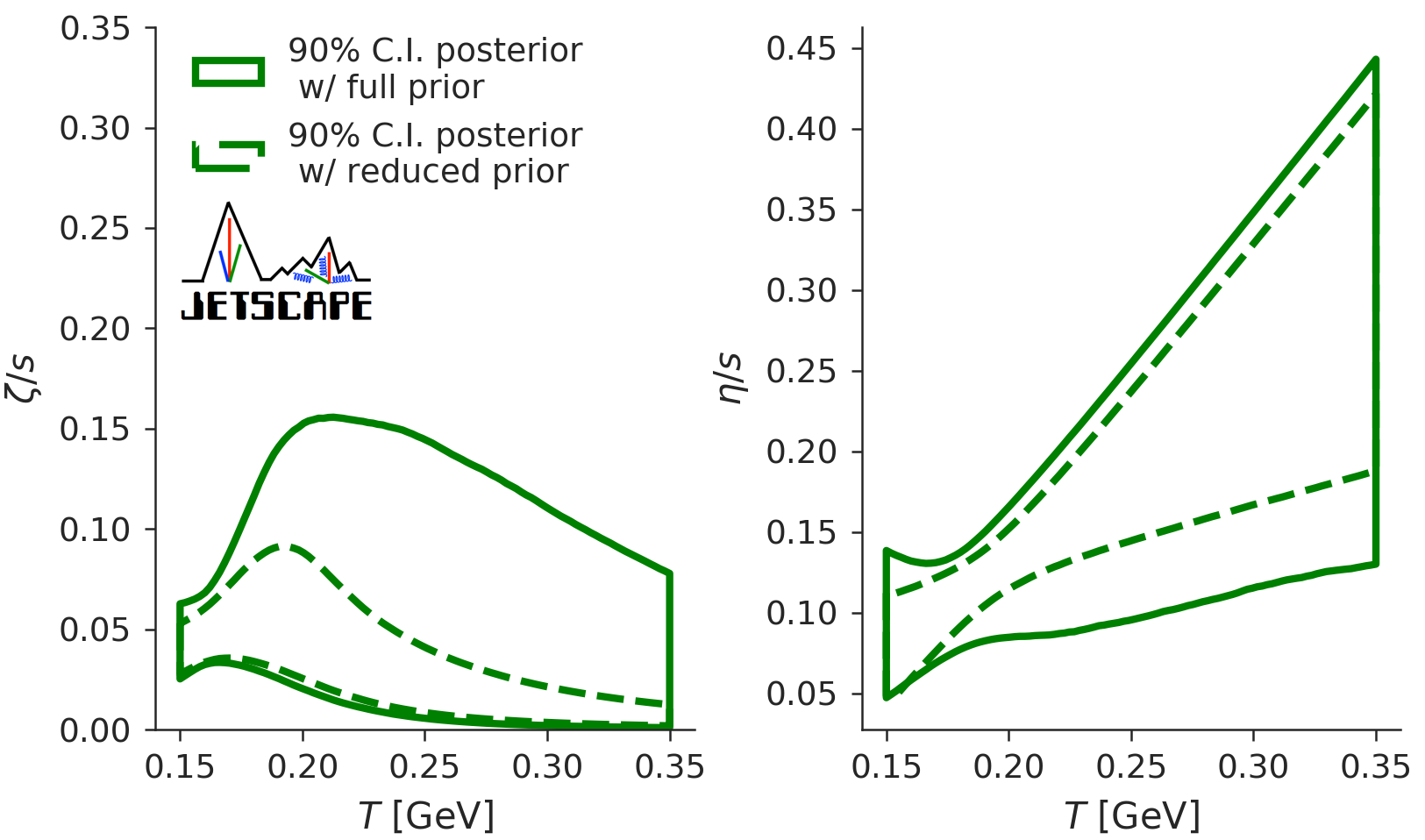}
					\caption{Constraints on shear and bulk viscosity obtained by comparing the same model of heavy-ion collisions with the same set of measurements, but reducing the parameter priors to match as closely as possible those of Ref.~\cite{Bernhard:2019bmu}. For the parametrization of shear and bulk viscosity, this reduction of the priors is equivalent to replacing the viscosity parametrization from the second row of Table~\ref{table:visc_param} by the functions in the first row (without curvature for $(\eta/s)(T)$ --- that is $\left(\eta/s\right)_{crv}=0$). 
					Figure reproduced from Ref.~\cite{JETSCAPE:2020mzn}; see reference for details.}
					\label{fig:effect_of_visc_param}
				\end{center}
			\end{figure}
			
			The parametrization of shear and bulk viscosity is an important modelling assumption. As seen in Table~\ref{table:visc_param}, the seven analyses used either of two parameterizations. The first group of analyses assumed that the specific shear viscosity $(\eta/s)(T)$ increased monotonically above a temperature of 154~MeV. Below this temperature, the viscosity was assumed to be a constant, with a discontinuity in $(\eta/s)(T)$ at this junction. The specific bulk viscosity was assumed to be a symmetric peak with a variable position, peak and width. The second group of analyses assumed that the specific shear viscosity $(\eta/s)(T)$ could be approximated by a piecewise linear function, with an inflection point at an adjustable temperature and variable slopes for $(\eta/s)(T)$ above and below this inflection point. This parametrization was continuous. For the specific bulk viscosity, the second group of analyses assumed an asymmetric peak with a variable position, peak value, width and asymmetry. An attempt to quantify the effect of using one parametrization or the other was made in Ref.~\cite{JETSCAPE:2020mzn}. It is possible to perform a Bayesian analysis restricting the second parametrization to the first (e.g., setting $\lambda_\zeta=0$), and also reducing the prior of the other model parameters to match closely the prior used in Ref.~\cite{Bernhard:2019bmu}. The result from Ref.~\cite{JETSCAPE:2020mzn} is shown in Figure~\ref{fig:effect_of_visc_param}. The effect of this reduction in the parameter priors is particularly large for bulk viscosity.
			
			The versions of relativistic viscous hydrodynamics used in the seven analyses, both DNMR and anisotropic hydrodynamics, have second-order transport coefficients. There are typically around a dozen of them, although there could be twice that number if some modelling assumptions were relaxed.
			In theory, all the second-order transport coefficients are functions of temperature, like shear and bulk viscosity are. From a pragmatic point of view, it seems unlikely that the temperature dependence of all these second-order coefficients can be constrained using flexible parameterizations for each one of them. At the moment, it is assumed that the second-order transport coefficients are related to the first-order ones and to thermodynamic quantities, using relations evaluated in simple versions of kinetic theory or holography. The effect of second-order transport coefficients was studied in Refs.~\cite{JETSCAPE:2020mzn,JETSCAPE:2020shq,Nijs:2020ors,Nijs:2020roc} by varying a constant multiplying the parametric forms obtained in kinetic theory or holography (see also Ref.~\cite{Liu-thesis} for a non-Bayesian study). These works found that the second-order coefficients had non-trivial correlations with other model parameters, implying that they do have a non-trivial effect on the result of the Bayesian inference. Understanding the effect of second-order transport coefficients remains a challenge for the field. On the other hand, as far as the results of the seven analyses under consideration are concerned, similar modelling assumptions were made by at least the six analyses using DNMR hydrodynamics, and it does not seem that second-order transport coefficients are a significant source of differences between the analyses.
			
			\paragraph{Cooper-Frye and post-hydrodynamics}
			
			As the temperature of the plasma decreases, quarks and gluons recombine into hadrons, and eventually the hydrodynamic description of the plasma breaks down. The exact temperature where this occurs is uncertain, but when it does, one must convert the energy-momentum tensor of hydrodynamics into matching hadronic momentum distributions for all species of hadrons, a process sometimes referred to as ``particlization''. Once hadrons are produced, their evolution is described with a relativistic hadronic transport such as UrQMD~\cite{Bass:1998ca,Bleicher:1999xi} or SMASH~\cite{Weil:2016zrk}. The matching between hydrodynamics and a kinetic description of hadronic is performed with the so-called ``Cooper-Frye'' prescription~\cite{Cooper:1974mv}. We discuss here three elements of the Cooper-Frye that have meaningful differences between the seven analyses of Table~\ref{table:bayes_viscosity}.
			
			The first one is related to the equation of state and the matching with the subsequent hadronic transport. As discussed above, the inclusion of the $\sigma$ meson in the hadron resonance gas modifies the equation of state used by four of the analyses from Table~\ref{table:bayes_viscosity}. A second consequence is that $\sigma$ mesons were produced by Cooper-Frye and decayed into pions, as explained in Ref.~\cite{Bernhard:2018hnz}. This increases the number of pions produced by the model. A discussion of the $\sigma$ meson can be found in Ref.~\cite{JETSCAPE:2020mzn}, although the overall consequence of this modelling choice on the Bayesian analysis has not been quantified.
			
			A second model difference in the treatment of Cooper-Frye is the sampling of the mass of the hadrons. Some analyses used the pole mass, while others sampled the mass from a Breit-Wigner distribution based on the original modelling choice of Refs~\cite{Bernhard:2019bmu,Bernhard:2018hnz}. Some discussion of this effect can also be found in Ref.~\cite{JETSCAPE:2020mzn}.
			
			A third difference is the model used to map the viscous part of the energy-momentum tensor of hydrodynamics to the hadronic momentum distribution, a modelling element often called ``viscous corrections''. In the absence of viscosity, it is assumed that the momentum distribution of hadrons is Bose-Einstein for bosons and Fermi-Dirac for fermions. When the effect of viscosity is non-negligible, different models have been used, based on an ansatz for the momentum distribution~\cite{Pratt:2010jt, Bernhard:2018hnz,McNelis:2019auj}, or approximate solutions of relativistic kinetic theory using the Grad expansion~\cite{Israel:1976tn, Israel:1979wp, Teaney:2003kp, Dusling:2009df, Monnai:2009ad, Dusling:2011fd, Denicol:2012cn} or the Chapman-Enskog expansion~\cite{chapman1990mathematical, 
				ANDERSON1974466, Jaiswal:2014isa}. Five of the analyses use the ansatz approach (either P.T.B.~\cite{Pratt:2010jt, Bernhard:2018hnz}, P.T.M.~\cite{Pratt:2010jt} or P.T.M.A.~\cite{McNelis:2019auj}). Heffernan et al studied Grad and Chapman-Enskog, while the JETSCAPE analysis studied four options side-by-side (Grad, Chapman-Enskog, P.T.B. and P.T.M.). The JETSCAPE and Heffernan analyses provide valuable insights into the effect of this modelling choice. The JETSCAPE study found significant differences in the parameter posteriors when different viscous corrections were used. Using Bayes factors, it also found that the Grad and P.T.B. viscous corrections were strongly favoured by data compared to the Chapman-Enskog one. Interestingly, Heffernan et al, which did use a completely different pre-hydrodynamics model,  found that Grad and Chapman-Enskog were equally favoured by the data (similar Bayes evidence). It does seem one should be careful before concluding that one viscous correction should be strongly favoured over another. On the other hand, it seems clear that different viscous corrections should be investigated, and that constraints on model parameters, including the viscosities, are unlikely to reach a satisfactory accuracy without accounting for this uncertainty.

			\paragraph{Choice of measurements}
			
			Heavy-ion physics is undoubtedly a data-rich field for Bayesian inference: there is a large number of measurements, and many of them have small uncertainties.  The measurements used in the seven analyses are a subset of the available ones. All analyses used boost-invariant hydrodynamics and only compared with midrapidity data sets. Two analyses, Moreland et al and Nijs et al, looked at ``small systems'', proton-lead collisions. The other analyses focused on collisions of large nuclei. Only the JETSCAPE analysis included measurements from lower collision energies at the Relativistic Heavy Ion Collider, while all other analyses focused solely on higher-energy Large Hadron Collider data sets (which JETSCAPE also included). Overall, few analyses used the same collision systems.
			
			For these various collision systems, four analyses used the same and similar data sets, which had only transverse-momentum integrated measurements. Three analyses added measurements: Nijs et al added transverse-momentum differential measurements (integrated observables are the first and sometimes second moments of the differential observables), while Heffernan et al and Parkkila et al added a partly overlapping sets of momentum anisotropy and event-plane correlation observables, with the latter also including symmetric cumulants.
			
			Nijs et al~\cite{Nijs:2020ors} performed a systematic comparison of Bayesian inference using different collision systems and subsets of measurements. JETSCAPE~\cite{JETSCAPE:2020mzn} compared the results obtained with Relativistic Heavy Ion Collider and Large Hadron Collider measurements separately. 
			
			\begin{figure}[tb]
				\begin{center}
					\includegraphics[width=0.7\textwidth]{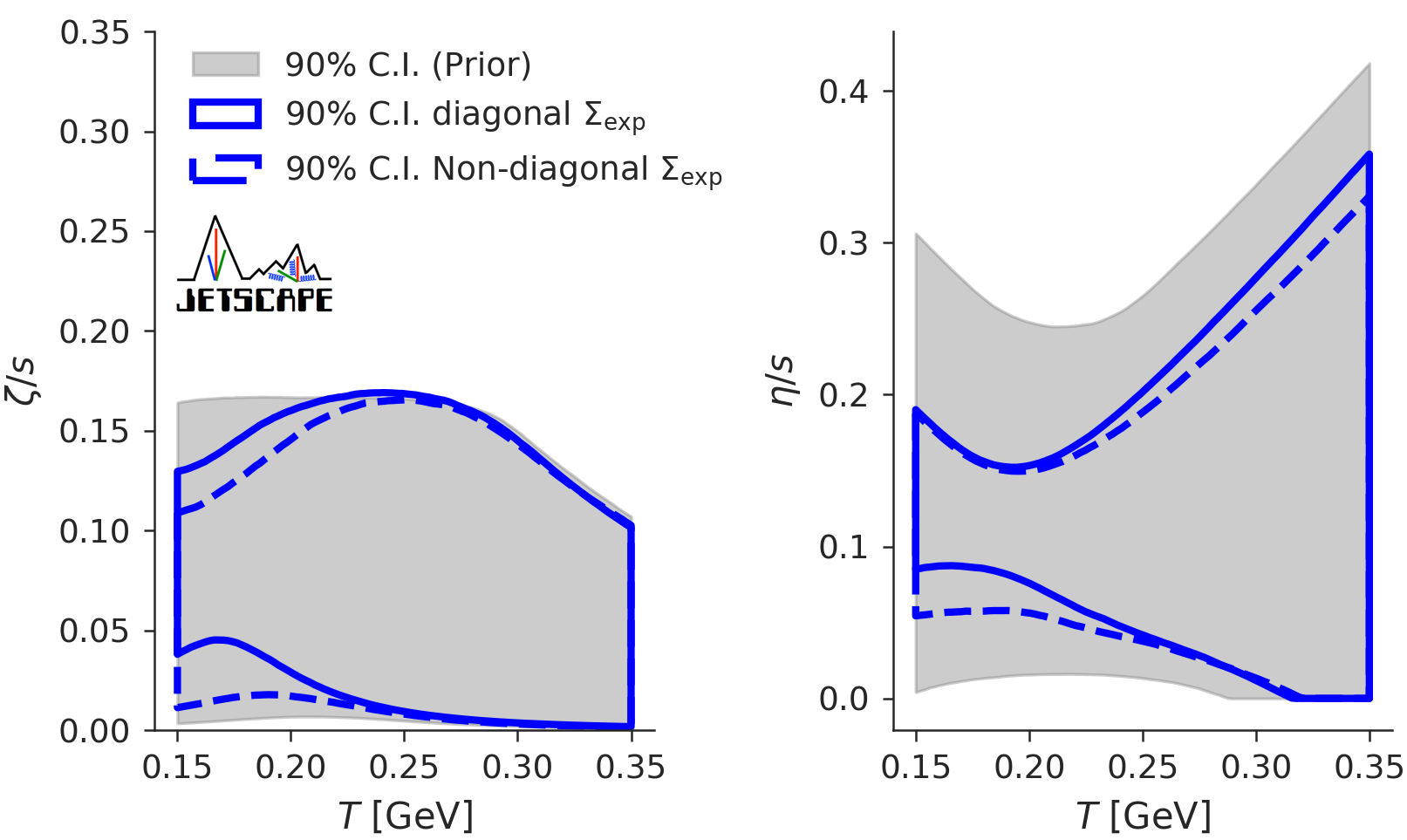}
					\caption{Effect of introducing correlations in the experimental covariance matrix between centralities and groups of observables that are expected to be correlated (see Ref.~\cite{JETSCAPE:2020mzn}
					for details). Both the statistical and systematic experimental uncertainties have the same correlations in this test, unlike in Ref.~\cite{Bernhard:2018hnz} and subsequent studies. Figure reproduced from Ref.~\cite{JETSCAPE:2020mzn}.}
					\label{fig:expt_covariance}
				\end{center}
			\end{figure}
			
			It is clear that analyses should strive to compare with large data sets. Whether the inclusion of more data sets necessarily leads to better constraints is a more subtle question. A perfect model of heavy-ion collisions should be compared with all possible data sets. However, an imperfect model may obtain better constraints on the parameters when comparing to a more limited data set. Small systems are a good example. The applicability of hydrodynamics should not be taken for granted even in collisions of large systems, but there is evidence from e.g., calculations of the Reynolds and Knudsen numbers~\cite{Niemi:2014wta}, that hydrodynamics might be a less accurate description of small systems such as proton-nucleus collisions. If that is the case, including measurements from proton-nucleus collisions in a Bayesian inference could bias the analysis by forcing it to describe data that it is not fully equipped to.
			
			That is not to say that one should not compare with these data sets. An important way to learn more about the model's applicability to e.g., small systems is to compare with these data sets. It can also be valuable to compare with these data sets in a Bayesian inference, to evaluate the model's ability to describe different data sets as systematically as possible. However, learning about the model's limitations is not the same objective as obtaining the best possible constraints on the model parameters, and it is possible that the latter is achieved by comparing with a limited data set.
			
			\paragraph{Covariance matrix}
			An extra challenge of the measurements is the lack of information about the experimental covariance matrix. How are uncertainties correlated across observables, across momentum, or across centralities? Some information is available, but only a small fraction of what would be needed. Bernhard et al addressed this by building a covariance matrix for the systematic experimental uncertainties using assumptions about the expected correlations of observables; many of the other analyses followed suit. The statistical uncertainties were assumed uncorrelated, although some correlations should be introduced as mentioned in Section~\ref{sec:likelihood}. A study of the effect of the covariance matrix was done in Ref.~\cite{JETSCAPE:2020mzn} and is shown in Figure~\ref{fig:expt_covariance}. Note that this test assumed that both the statistical and systematic experimental uncertainties had the same correlations, unlike the studies listed in Table~\ref{table:bayes_viscosity}. It is clear that the assumptions about the experimental covariance matrix have an effect on the posterior, in this case, on the shear and bulk viscosity constrained from data.
			
			Two analyses, Bernhard et al and Parkkila et al, included an additional source of uncertainty in their covariance matrix: the $\Sigma_{extra}$ from Table~\ref{table:bayes_viscosity}. This additional uncertainty was meant to account for imperfections in the model. The details of the implementation can be found in Ref.~\cite{Bernhard:2018hnz}. The presence of this uncertainty in some but not all the analyses makes their results difficult to compare. 
			In general, a better-controlled alternative to the addition of this $\Sigma_{extra}$ is to relax modeling assumptions. 
			For example, there are several reasonable models providing initial conditions for the hydrodynamics equations, and it is better to perform a Bayesian analysis with multiple initial condition models, rather than to use a single model and include an additional uncertainty like $\Sigma_{extra}$ to attempt to account for the missing ``initial condition uncertainty''.

			\paragraph{Assessment}
			
			The seven Bayesian analyses listed in Table~\ref{table:bayes_viscosity} made different modelling choices, compared with different data sets, and obtained different constraints on the model parameters, including the shear and the bulk viscosities. In general, more data increases precision, while improved modelling provides increased accuracy. The analyses that currently provide the most accurate constraints on the viscosities is likely the one making the best modelling choices. Which are the best modelling choices is a question that reasonable people would disagree on. However, there are clear lessons to be learned from the exercise:
			\begin{itemize}
				\item It is valuable to perform Bayesian inference with different modelling choices (e.g., Trento vs IP-Glasma, DNMR vs anisotropic hydrodynamics, viscous corrections, ....) to better understand how robust are constraints on the parameters; 
				\item There is a number of reasons why analyses by different groups obtain different constraints on model parameters, and these differences can generally be understood systematically; manuscripts must share enough information such that the modelling choices of each analysis are clear;
				\item Adding more measurements can improve precision, but not necessarily accuracy; it is essential to constantly revisit the modelling choices, since accuracy is likely limited by approximations in the physics of the model; on the other hand, comparing with larger data set is a useful way to identify areas of the model that need improvements;\footnote{If adding measurements reduce precision, it can be an important indication that the model struggles to describe all measurements simultaneously.}
				\item Statistical and interpolation uncertainties from the emulators were not discussed in detail because they can be challenging to compare across analyses. However, there is no doubt that they differed in the studies from Table~\ref{table:bayes_viscosity}, and it is very unlikely that the emulator uncertainties are significantly smaller than the experimental ones. It is thus important to remember that posteriors from Bayesian inference often do not make full use of the current accuracy of the experimental data (they contain numerical uncertainties).
				\item In current analyses, significant assumptions are made regarding the experimental covariance matrix: correlations between observables are neglected or are estimated using theoretical input. It may be unrealistic to expect to ever have access to the complete experimental covariance matrix with information on the correlation between every  observable. Yet it would already be very helpful to have information such as the \emph{correlation length} of uncertainties across centrality, rapidity or transverse momentum bins. For example, are uncertainties as strongly correlated between the lowest and highest rapidity bin as they are for two neighbouring rapidity bins? Moreover, if multiple observables are studied within an experimental analysis, correlation of uncertainties across different observables should be calculable, and publishing this information would provide important inputs for Bayesian analyses.
			\end{itemize}

			\section{Future of Bayesian inference}
			
			\label{sec:future}

			\subsection{Experimental design}
			
			\label{sec:expt_design}

			Closure tests were discussed in Section~\ref{sec:closure_tests}. It is the process of performing Bayesian inference on model calculations, to help validate the emulator and the entire Bayesian process before comparison with data. There is no reason to limit closure tests to measured observables: from the output of numerical simulations of heavy-ion collisions, any observable can be calculated, and one can include any subset of these observables in closure tests. By doing so, one can help establish, before any measurement is performed, how much constraints on various model parameters will be provided by a given observable. Note that this is far beyond verifying if an observable is sensitive to a given parameter. By including a new observable in the Bayesian inference, one can:
			\begin{itemize}
				\item Verify if the new observable provides any additional constraints beyond what is already provided by other observables; 
				\item Determine the relation between the experimental uncertainty on the measurement and the constraints on the model parameters;
				\item Establish if the new observable can distinguish between competing models.
			\end{itemize}
			
			We use ``experimental design'' as an umbrella term for the idea of guiding new measurements, or even the entire design of an experiment, based on a statistical analysis of a model's current predictions. Ref.~\cite{Sangaline:2015isa} provided the first large-scale attempt at experimental design to link numerical simulations of heavy-ion collisions and constraints provided by measurements. A more recent work, Ref.~\cite{Nijs:2021clz}, looked specifically at the example of planned oxygen-oxygen collisions at the Large Hadron Collider, and how these measurements could help constrain different model parameters. 
			
			A more general discussion of experimental design as applied to nuclear physics can be found in Ref.~\cite{Phillips:2020dmw}. It emphasizes broader aspects of the topic, in particular 
			the power of attempting to quantify the benefits and costs of measurements into a ``utility function'', which can then be optimized to find measurements that lead to the highest return on investment.
			
			Overall, there are many more opportunities with experimental design than have been explored yet. It is clear that we should use this tool more often to provide guidance on new experimental measurements. 
			There are caveats, however. Experimental design is model dependent, just like Bayesian inference in general. Importantly, the statistical and interpolation uncertainties in the emulators can skew the results of experimental design if these uncertainties are not much smaller than the experimental uncertainty: halving the experimental uncertainty of a given observable could artificially show no effect in an experimental design test if the emulator uncertainty is large for this observable. Therefore, it is important to remember the non-trivial modelling biases of experimental design tests.

			\subsection{Modeling improvements and uncertainties}
			
			\label{sec:model_uncert_future}

			\begin{figure}[h]
				\centering
				\begin{minipage}[b]{0.3\textwidth}
					\centering
						\begin{tikzpicture}
						\begin{axis}[
							domain=-3:3,
							samples=100,
							axis lines=left,
							xlabel=Parameter,
							ylabel=Posterior,
							height=6cm,
							width=6cm,
							xticklabels={,,},
							yticklabels={,,},
							ticklabel style={font=\tiny},
							]
							\addplot [very thick, smooth,blue!50!black] {exp(-(x-.9)^2/2/.06) / (sqrt(2*pi)*1)};
							\draw [line width=0.6mm,black,dashed] (axis cs:0.0,0.0) -- (axis cs:0.0,0.6);
						\end{axis}
						\node[above] at (current bounding box.north) {Precise \& inaccurate};
					\end{tikzpicture}
					\label{fig:precise_inaccurate}
				\end{minipage}
				\hfill
				\begin{minipage}[b]{0.3\textwidth}
					\centering
					\begin{tikzpicture}
					\begin{axis}[
						domain=-3:3,
						samples=100,
						axis lines=left,
						xlabel=Parameter,
						ylabel=Posterior,
						height=6cm,
						width=6cm,
						xticklabels={,,},
						yticklabels={,,},
						ticklabel style={font=\tiny},
						]
						\addplot [very thick,smooth,blue!50!black] {exp(-x^2/2/2.) / (sqrt(2*pi)*1)};
						\draw [line width=0.6mm,black,dashed] (axis cs:0.0,0.0) -- (axis cs:0.0,0.6);
					\end{axis}
											\node[above] at (current bounding box.north) {Imprecise \& accurate};
					\end{tikzpicture}
					\label{fig:accurate_imprecise}
				\end{minipage}
				\hfill
				\begin{minipage}[b]{0.3\textwidth}
					\centering
					\begin{tikzpicture}
						\begin{axis}[
							domain=-3:3,
							samples=100,
							axis lines=left,
							xlabel=Parameter,
							ylabel=Posterior,
							height=6cm,
							width=6cm,
							xticklabels={,,},
							yticklabels={,,},
							ticklabel style={font=\tiny},
							]
							\addplot [very thick,smooth,blue!50!black] {exp(-x^2/2/.1) / (sqrt(2*pi)*1)};
							\draw [line width=0.6mm,black,dashed] (axis cs:0.0,0.0) -- (axis cs:0.0,0.6);
						\end{axis}
						\node[above] at (current bounding box.north) {Precise \& accurate};
					\end{tikzpicture}
					\label{fig:precise_accurate}
				\end{minipage}
				\caption{\label{fig:precision_vs_accuracy}Schematic comparison of a one-parameter posterior (solid blue line) with the hypothetical true value of a parameter (dashed black line). (Left) Precise but inaccurate; (Center) Accurate but imprecise; (Right) Accurate and precise. }
			\end{figure}
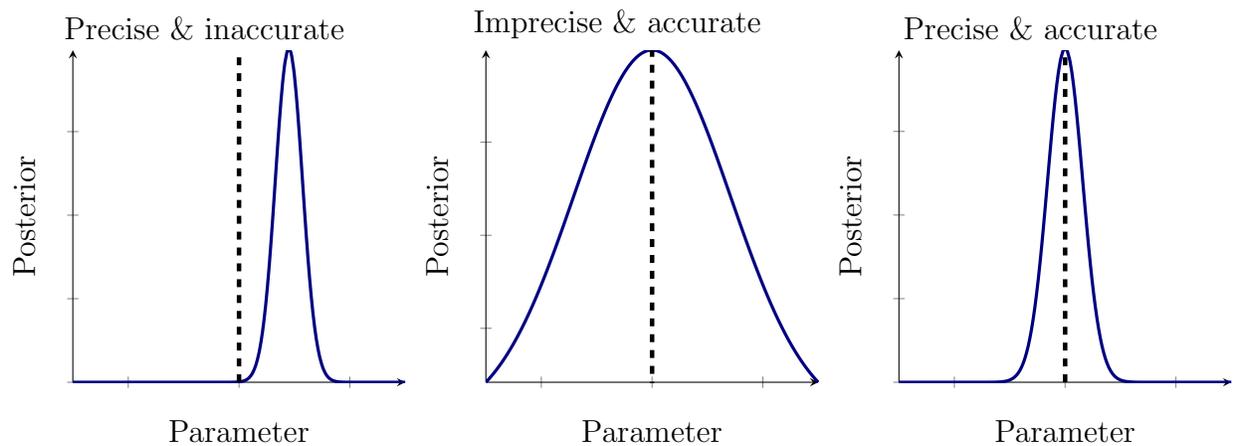

			Another essential ingredient in obtaining more reliable constraints on model parameters is improving the physics description of the collisions, which are extremely complex to model. While many features of the data can be associated with specific phenomena --- such as the relation between the momentum anisotropy of final state particles and the initial geometry of the collision --- a precise quantitative description of broad sets of measurements requires a comprehensive numerical model. Modern multistage simulations of heavy-ion collisions take into account a wide range of effects that have been found to be important to describing measurements, yet it is evidently understood that the models are not perfect. Comparing a model with more and more measurements without taking into account the imperfection of the model can lead to very precise constraints that are inaccurate (Figure~\ref{fig:precision_vs_accuracy} left). Focusing on quantifying model uncertainties with insufficient measurements can yield accurate constraints that will not be very constraining (Figure~\ref{fig:precision_vs_accuracy} center). 
			
			Improving models generally requires relaxing assumptions. Unless there is strong theoretical evidence for e.g., an initial condition model or a choice of transport coefficient parametrization, Bayesian inference should be performed with different choices of models. Two obstacles to this can be readily addressed. The first one is the practical difficulty of using a new model because the computer code necessary for it is not available. This can be solved by a focus of the community on making research codes available publicly (see the next section). Second, while performing a Bayesian inference with multiple variations of a model can be numerically demanding, tools such as transfer learning (Section~\ref{sec:emulation_new_devs}) can make this possible at a significantly lower computational cost.
			
			Statistical tools can also help with this task. For example, Bayesian model mixing discussed in Section~\ref{sec:model_averaging} can provide a systematic approach to combine the strength of different models, to help build a better model from imperfect ones.

			\subsection{Open science and data management}
			
			\label{sec:data_management}
			
			The current level of progress with Bayesian inference in heavy-ion collisions would not have been possible without the efforts of multiple groups to document and release computer codes that simulate heavy-ion collisions. Bayesian inference is inevitably a numerical effort in heavy-ion collisions. While much can be studied from simplified models, the idea of Bayesian inference is to use state-of-the-art models to describe simultaneously as many measurements as possible. One needs easy access to all models of heavy-ion collisions that are relevant to compare with measurements, whether these are initial condition models, different versions of hydrodynamics, equations of state, etc. Modular codes are particularly important since one may want to replace a specific part of the model without changing the rest. Multiple numerical simulations of heavy-ion collisions are now available, including iEBE~\cite{iebe_code} and iEBE-MUSIC~\cite{iebe_music_code}, the JETSCAPE 2D soft-sector package~\cite{sims_2d_code,sims_2d_jf_code} and the JETSCAPE~\cite{jetscape_code} and XSCAPE frameworks~\cite{xscape_code},  Trajectum~\cite{trajectum_code}, and many others, all building on efforts of dozens of members of the community to make their individual simulations publicly available.

			Codes to perform Bayesian inference are also available publicly~\cite{param_est_code,sims_param_est_code,stat_param_est_code,dan_ahydro_bayes_code,band_code}, many based on the code written for Ref.~\cite{Bernhard:2019bmu} as well as from recent efforts by the JETSCAPE~\cite{jetscape_website} and BAND Collaborations~\cite{Phillips:2020dmw}.
			
			\paragraph{Sharing emulators and calculations}
			An important next step is to find ways to share calculations. A very large amount of computational time on supercomputers is required to obtain the calculations necessary to perform Bayesian inference. Once model calculations are performed, the training of the emulators and the Bayesian inference are relatively inexpensive. There is every reason to make public either the raw calculations or the emulators. It allows other groups to reuse the model calculations, which can be educational but can also allow other groups to build on published results and go beyond. An option to share results is to share the emulators in a format suitable for long-term storage\footnote{Some groups have used Python ``pickle'' and ``dill'' objects for emulator storage, but these have not proved successful for long-term storage.}, and with documentation on how to use the emulators to obtain observables given a set of model parameters.
			
			As far as sharing calculations, an important step would be to agree on formats for the observables. Work has been done by different groups, including the JETSCAPE Collaboration~\cite{sims_param_est_code}, to identify an efficient data-reduction approach to save sufficient information to reconstruct most low-energy hadron observables without saving the entire particle shower. Community-agreed formats would help data management and sharing. A computer library developed and maintained by the community (theory and experiment), to calculate observables from the list of final particles, would play an important role, and would accelerate the inclusion of new observables in Bayesian analyses. Such a computer library could build on existing efforts~\cite{Bailey:2022tdz}, such as RIVET~\cite{Bierlich:2020wms,Buckley:2010ar}, though potentially optimized for the computing needs of large-scale analyses, which often require aggressive data reduction such that only a small amount of information is saved for each collision.

			\section{Summary and outlook}
			
			\label{sec:summary}
			
			Bayesian inference provides a powerful probabilistic framework to compare heavy-ion measurements with model calculations. One of its greatest strengths is its ability to make full use of the experimental uncertainties of measurements, combining them with all other uncertainties (model, numerical, statistical) into probabilistic constraints on the model parameters. Another strength of the approach is to force practitioners to quantify and address the limitations of our models of heavy-ion collisions, and provides tools to do it.
			
			We discussed the role of emulation in Section~\ref{sec:emulation}, and the focus on Gaussian process emulators for their ability to quantify both the emulator's prediction and non-negligible predictive uncertainty. The necessity of quantifying accurately the emulator's interpolation uncertainty is an important criterion to consider when investigating machine learning substitutes for Gaussian processes.
				
			In Section~\ref{sec:bayes_applications}, we discussed seven different analyses that looked at somewhat similar measurements with generally similar models, and obtained different constraints on the parameters. What is clear is that applications of Bayesian inference in heavy-ion collisions are a work in progress. The field is healthy, with different groups performing analyses and attempting to verify each other's work. Importantly, there has been a sustained effort to perform these studies openly, providing access to numerical codes so that progress can occur faster and more transparently. 
			
			Because Bayesian inference provides quantified uncertainty bands, it is natural to expect that these bands are an accurate reflection of the uncertainties in the data and the calculations. As discussed in Section~\ref{sec:bayes_applications}, the reality is more nuanced: simulations of heavy-ion collisions are complex, and known sources of modelling uncertainties cannot always be included comprehensively. We also saw with Figure~\ref{fig:effect_of_visc_param} that decisions such as the details of the parametrization of bulk viscosity, which may seem relatively harmless, can have large effects on the results of the analysis. This knowledge is being used, along with other advances, to shape the next generation of Bayesian studies of heavy-ion data. With a sustained community effort and a focus on open science, we can expect to rapidly improve constraints on the properties of the quark-gluon plasma.

			\ack
			I am indebted to Dananjaya Liyanage and Matthew Heffernan for their invaluable feedback on this work, and to Matthew Luzum for his insights regarding uncertainty correlations.			
			Large parts of this review are built on discussions and collaborations with my former colleagues at Duke, and with current and former members of the JETSCAPE Collaboration, many of whom wrote outstanding Ph.D. theses that are cited throughout this review. 
			I thank Steffen Bass, Jonah Bernhard, Yi Chen, Jacob Coleman, Lipei Du, Raymond Ehlers, Derek Everett, Wenkai Fan, Charles Gale, Matthew Heffernan, Ulrich Heinz, Peter Jacobs, Yi Ji, Weiyao Ke, Dananjaya Liyanage, Matthew Luzum, Abhijit Majumder, Simon Mak, Andi Mankolli, James Mulligan, Govert Nijs, Scott Moreland, Scott Pratt, Chun Shen, Ron Soltz, Wilke van der Schee, Gojko Vujanovic and Yingru Xu for discussions and collaborations. I thank Jasper Parkkila for his feedback on the manuscript. This work was supported in part by the U.S. Department of Energy, Office of Science under Award Numbers DE-FG-02-05ER41367  and DE-SC-0024347.
			\clearpage
			
			\appendix
			
			\bibliographystyle{iopart-num}
			\bibliography{biblio_inspireshep,biblio_others}
			
		\end{document}